\documentclass[a4paper,twocolumn,superscriptaddress]{revtex4-1}
\usepackage{amsmath}
\usepackage{amssymb}
\usepackage{amsfonts}
\usepackage{graphicx}
\usepackage{color}
\usepackage{esint}
\usepackage{bm}

\usepackage[colorlinks=true,citecolor=magenta,linkcolor=magenta]{hyperref}
\usepackage{subfigure}

\begin{document}

\title{Bump formation in the runaway electron tail}

\author{J. Decker}
\email{joan.decker@epfl.ch}
\affiliation{Ecole Polytechnique F\'{e}d\'{e}rale de Lausanne (EPFL), Centre de Recherches\\
 en Physique des Plasmas (CRPP), CH-1015 Lausanne, Switzerland}

\author{E. Hirvijoki}
\affiliation{Department of Applied Physics, Chalmers University of Technology,
SE-41296 Gothenburg, Sweden}

\author{O. Embreus}
\affiliation{Department of Applied Physics, Chalmers University of Technology,
SE-41296 Gothenburg, Sweden}

\author{Y. Peysson}
\affiliation{CEA, IRFM, F-13108 Saint-Paul-lez-Durance, France}

\author{A. Stahl}
\affiliation{Department of Applied Physics, Chalmers University of Technology,
SE-41296 Gothenburg, Sweden}

\author{I. Pusztai}
\affiliation{Department of Applied Physics, Chalmers University of Technology,
SE-41296 Gothenburg, Sweden}

\author{T. F\"{u}l\"{o}p}
\affiliation{Department of Applied Physics, Chalmers University of Technology,
SE-41296 Gothenburg, Sweden}

\date{\today}
\begin{abstract}
Runaway electrons are generated in a magnetized plasma when the
parallel electric field exceeds a critical value. For such electrons
with energies typically reaching tens of MeV, the
Abraham-Lorentz-Dirac (ALD) radiation force, in reaction to the
synchrotron emission, is significant and can be the dominant process
limiting the electron acceleration. The effect of the ALD-force on
runaway electron dynamics in a homogeneous plasma is investigated
using the relativistic finite-difference Fokker-Planck codes LUKE
[Decker \& Peysson, Report EUR-CEA-FC-1736, Euratom-CEA, (2004)] and CODE
[Landreman et al, Comp.~Phys.~Comm.~{\bf 185}, 847 (2014)]. Under the action 
of the ALD force, we find that a bump is formed in the tail of
the electron distribution function if the electric field is sufficiently
large. We also observe that the energy of runaway electrons in the bump
increases with the electric field amplitude, while the population
increases with the bulk electron temperature. The presence of the bump
divides the electron distribution into a runaway beam and a bulk
population. This mechanism may give rise to beam-plasma types of
instabilities that could in turn pump energy from runaway electrons
and alter their confinement.
\end{abstract}
\maketitle

\section{Introduction}

\label{sec:intro} 
Runaway electrons are typically generated in plasmas in the presence
of large electric fields $E>E_{c}$, where the critical field
$E_{c}$ is defined as \cite{dre59}
\begin{equation}
E_{c}=\frac{ne^{3}\ln\Lambda}{4\pi\varepsilon_{0}^{2}mc^{2}},
\end{equation}
where $n$ is the electron density, $m$ is the electron rest mass,
$c$ is the speed of light, $e$ is the elementary charge, and $\ln\Lambda$
is the Coulomb logarithm.

In connection with the sudden cooling in a tokamak disruption, a strong
electric field is induced, which leads to the generation of a large
number of runaway electrons. In certain cases a significant fraction of
the initial toroidal current can be driven by a beam of
runaway electrons. The
formation of such energetic runaway beams would represent a serious
threat for reactor-size machines such as ITER\cite{hen07}.
Consequently, a considerable research effort is currently undertaken 
to prevent the formation of large runaway beams during tokamak
disruptions, or to design a controlled damping scenario for runaway
beams if they cannot be avoided \cite{hen07,izz11,leh11,gra14,hol15}.
The condition $E>E_{c}$ for runaway electron generation
can also be met during the plasma start-up or ramp-down. During the flat-top
phase, runaways can appear if the density is sufficiently low in Ohmic plasmas (since $E_{c}\propto n$), or if an
externally applied source of current is suddenly modified. 

Experimental measurements show that the maximum runaway electron
energy does not increase indefinitely with time but instead reaches
a limit in the tens of MeV range \cite{hol13}. One of the possible mechanisms
that could provide an explanation for this limit is the
Abraham-Lorentz-Dirac (ALD) radiation force \cite{pau58} in reaction
to the synchrotron emission due to the particle motion in a magnetic
field. For electrons in the
MeV range, this force can be significant and contribute to limit the
particle acceleration \cite{and01}. The synchrotron emission, which
carries energy away from the electrons, is also used as a diagnostic
tool for the runaway population \cite{sta13}. The ALD force 
is characterized by the synchrotron radiation reaction time scale $\tau_r$, given by
\begin{equation}
\tau_{r}^{-1}=\frac{e^{4}B^{2}}{6\pi\varepsilon_{0}(mc)^{3}},\label{eq:taur}
\end{equation}
where $B$ is the magnetic field.

In the present paper, the runaway electron dynamics in a homogeneous
plasma is investigated using the relativistic finite-difference
guiding-center Fokker-Planck codes LUKE \cite{dec04a,pey14} and CODE
\cite{lan14,sta15}. The electron distribution function evolves under the
combined influence of Coulomb collisions, electric field acceleration,
and the ALD radiation reaction force. Under a constant parallel
electric field (with respect to the magnetic field)
$E_{\Vert}>E_{c}$, the electron distribution never reaches a
steady-state in the absence of ALD radiation reaction force. Conversely, it is
shown in Sec.~\ref{sec:evolution} that when the effect of the ALD force is included, the electron
distribution evolves towards a steady-state solution. This solution
exists even though the synchrotron emission vanishes for electrons
with purely parallel motion in a uniform magnetic field. In fact, the
expansion of the electron distribution towards higher energies is
limited by collisional pitch-angle scattering, which is enhanced by
the strong perpendicular anisotropy arising from the combination of
electric field acceleration and synchrotron radiation reaction
force. This process is found to limit the runaway electron population
to energies far below the value for which the contribution from the
magnetic field curvature to the ALD radiation reaction force becomes
significant \cite{and01}. Consequently, it is justified to use the
homogeneous plasma limit to study the dynamics of runaway electrons in
the core region of tokamaks.

In addition, we show that if the electric field amplitude is sufficiently
large, a bump appears in the runaway electron tail of the steady-state
distribution function, in accordance with analytical predictions \cite{hir15}.
This bump, which peaks on the parallel axis in momentum space, is
entirely located in the runaway region. The steady-state population
of electrons in the bump is found to increase with the bulk electron
temperature $T_{e}$, while their average energy increases with the
electric field amplitude $E_{\Vert}$ and decreases with the amplitude
of the ALD radiation reaction force, which is proportional to $B^{2}$.
For certain parameters, the bump in the electron distribution tail encompasses almost
the entire runaway electron population, thus formally dividing the
distribution into a bulk population and a runaway beam.

The implementation of the synchrotron reaction force in the kinetic
equation is described in Sec.~\ref{sec:kinetic_equation}. The time
evolution of the electron distribution calculated by the Fokker-Planck
modelling code LUKE is presented in Sec.~\ref{sec:evolution}. The properties of the steady-state
distribution function and the mechanism leading to the formation of a bump
are described in Sec.~\ref{sec:Properties}, where a comparison between the
codes LUKE and CODE is also presented. The bump is
characterized in Sec.~\ref{sec:Parametric} as a function of the
electric field amplitude, ALD radiation reaction, bulk electron
temperature, and ion effective charge. Implications of the ALD
radiation reaction force and bump-in-tail formation are discussed in
the Conclusions, Sec.~\ref{sec:Conclusions}.

\section{Synchrotron reaction force in the kinetic equation\label{sec:kinetic_equation} }

\subsection{Kinetic equation for charged particles in a magnetized plasma}

The kinetic equation for species $a$ with charge $q$ and mass $m$ is given by 
\begin{align}
\frac{\partial f_{a}}{\partial t}+\frac{\partial}{\partial\mathbf{x}}\cdot\left(\dot{\mathbf{x}}f_{a}\right)+\frac{\partial}{\partial\mathbf{p}}\cdot\left(\dot{\mathbf{p}}f_{a}\right)=C[f_{a},f_{b}],\label{eq:kinetic}
\end{align}
where $C[f_{a},f_{b}]$ is the collision operator between particle
species $a$ and $b$ (including intra-species collisions) and $(\dot{\mathbf{x}},\dot{\mathbf{p}})$
are the equations of motion associated with phase-space coordinates
$(\mathbf{x},\mathbf{p})$. Here $\mathbf{x}$ is the particle position and
$\mathbf{p}=\gamma m\mathbf{v}$ is the particle momentum, with
$\gamma=1/\sqrt{1-v^{2}/c^{2}}=\sqrt{1+p^{2}/(mc)^{2}}$ the relativistic
factor. In the Fokker-Planck limit,
the Coulomb collision operator is given by 
\begin{align}
C_{\textrm{FP}}[f_{a},f_{b}]=-\frac{\partial}{\partial\mathbf{p}}\cdot\left(\mathbf{K}_{\textrm{FP}ab}[f_{b}]f_{a}-\mathbb{D}_{\textrm{FP}ab}[f_{b}]\cdot\frac{\partial f_{a}}{\partial\mathbf{p}}\right),
\end{align}
where $\mathbf{K}_{ab}[f_{b}]$ is the collisional friction vector
and $\mathbb{D}_{ab}[f_{b}]$ is the collisional diffusion tensor. 
The relativistic Braams-Karney collision operator
is used in this paper \cite{bra87,shk97}. 

So-called \textit{knock-on} collisions represent a $1/\ln\Lambda$
correction to the collision operators. However, when the runaway population
becomes significant, these collisions can play an important role as
they give rise to an avalanche effect that can significantly increase
the runaway growth rate. This secondary runaway generation is neglected in
the present work, which is restricted to situations where the runaway
population is sufficiently small for secondary electron generation
to be negligible. However, it is possible that the ALD radiation reaction force
has a significant effect on the secondary runaway generation. Such considerations
will be the subject of future work.

The equations of motion 
combine the Hamiltonian motion from the electric and magnetic fields
$\mathbf{E}$ and $\mathbf{B}$, and the effect of the ALD radiation
reaction $\mathbf{F}_{\textrm{ALD}}:$ 
\begin{align}
\dot{\mathbf{x}}= & \mathbf{v},\\
\dot{\mathbf{p}}= & q\left(\mathbf{E}+\mathbf{v}\times\mathbf{B}\right)+\mathbf{F}_{\textrm{ALD}}
\equiv\mathbf{F}_{E}+\mathbf{F}_{m}+\mathbf{F}_{\textrm{ALD}}.
\end{align}
The Abraham-Lorentz-Dirac force describes momentum loss in
reaction to the synchrotron radiation, and takes the form~\cite{pau58} 
\begin{align}
\mathbf{F}_{\textrm{ALD}}=&\frac{q^{2}\gamma^{2}}{6\pi\varepsilon_{0}c^{3}}\left[\ddot{\mathbf{v}}+\frac{3\gamma^{2}}{c^{2}}\left(\mathbf{v}\cdot\dot{\mathbf{v}}\right)\dot{\mathbf{v}}\right.\notag\\
&\left.+\frac{\gamma^{2}}{c^{2}}\left(\mathbf{v}\cdot\ddot{\mathbf{v}}+\frac{3\gamma^{2}}{c^{2}}\left(\mathbf{v}\cdot\dot{\mathbf{v}}\right)^{2}\right)\mathbf{v}\right].\label{eq:reaction_force}
\end{align}

In magnetically confined fusion plasmas, the magnetic force $\mathbf{F}_{m}=q\mathbf{v}\times\mathbf{B}$
characterized by the Larmor frequency $\omega_{c}=qB/m$ typically
dominates both the electric force $\mathbf{F}_{E}=q\mathbf{E}$ and
the radiation reaction force $\mathbf{F}_{\textrm{ALD}}$ such that
$\mathbf{v}\cdot\dot{\mathbf{v}}\simeq0$. In a uniform constant magnetic
field, the ALD force (\ref{eq:reaction_force}) thus reduces to 
\begin{equation}
\mathbf{F}_{\textrm{ALD}}\simeq-\frac{m}{\tau_{r}}\left[\mathbf{v}_{\bot}+\frac{\gamma^{2}v_{\bot}^{2}}{c^{2}}\mathbf{v}\right],\label{eq:K}
\end{equation}
where
$\mathbf{v}_{\bot}=\left(\mathbf{I}-\hat{\mathbf{b}}\hat{\mathbf{b}}\right)\cdot\mathbf{v}$
is the perpendicular velocity with norm $v_{\bot}=\left\Vert
\mathbf{v}_{\bot}\right\Vert $ and $\hat{\mathbf{b}}=\mathbf{B}/B$ is
the magnetic field unit vector.

\subsection{Guiding-center transformation \label{sub:gc}}

In fusion plasmas, the gyroperiod is short compared to the time scale
associated with collisions, the ALD radiation reaction force, and the
electric field acceleration. Based on this time-scale separation, the
kinetic equation is reduced by eliminating the gyromotion in
Eq.~(\ref{eq:kinetic}) using Lie-transform perturbation methods
\cite{bri04,dec10b}. The transformation of the dissipative ALD force
uses the Lie-transform for non-Hamiltonian dynamics, which has been
recently derived in Ref. \cite{hir14}.  In a uniform plasma, the resulting
guiding-center distribution function for electrons evolves in the 2-D
gyro-angle independent momentum space $(p,\xi)$ as

\begin{equation}
\frac{\partial f}{\partial t}+\boldsymbol{\nabla}_{p,\xi}\cdot\mathbf{S}_{p,\xi}\left[f\right]=I_{\textrm{FP}}[f],\label{eq:gc_kinetic}
\end{equation}
where $p$ is the guiding-center momentum and $\xi=p_{\parallel}/p$ is
the pitch-angle cosine. Components of the guiding-center
momentum-space flux
\begin{equation}
\mathbf{S}_{p,\xi}\left[f\right]=\left(\mathbf{K}_{\textrm{FP}}+\mathbf{K}_{\textrm{E}}+\mathbf{K}_{\textrm{ALD}}\right)f-\mathbb{D}_{\textrm{FP}}\cdot\boldsymbol{\nabla}_{p,\xi}f
\end{equation}
include convective contributions from collisional drag
$\mathbf{K}_{\textrm{FP}}$, electric field acceleration
$\mathbf{K}_{\textrm{E}}$, and the radiation reaction force
$\mathbf{K}_{\textrm{ALD}}$, and a collisional diffusion tensor
$\mathbb{D}_{\textrm{FP}}$. The integral part of the guiding-center
collisional operator is denoted $I_{\textrm{FP}}[f]$. It describes the
evolution of the bulk population due to collision with fast electrons.
When momentum conservation of the electron-electron collision operator
is essential - as for the calculation of electron-driven current - the
term $I_{\textrm{FP}}[f]$ must be included \cite{kar86}. It is generally truncated
at the first order in Legendre expansion. The truncation ensures
momentum conservation but allows energy dissipation such that it is
not necessary to model energy transport to reach a steady-state
solution. In the present paper, the term is set to zero unless
otherwise specified.  Omitting $I_{\textrm{FP}}[f]$ makes it possible
to use the stream function to interpret the steady-state fluxes in
momentum space, as seen in Sec.~\ref{sub:Steady-state-solution}. It
also allows a comparison between the codes LUKE and CODE, since the
integral part of the collision operator is not yet included in
CODE. This benchmark is presented in Sec.~\ref{sub:Benchmark},
where it is also shown that the effect of $I_{\textrm{FP}}[f]$ on the
tail of the electron distribution can be neglected.

Writing out the momentum space divergence operator explicitly, the kinetic equation (\ref{eq:gc_kinetic})
becomes
\begin{equation}
\frac{\partial f}{\partial t}+\frac{1}{p^{2}}\frac{\partial}{\partial p}\left(p^{2}S_{p}\right)-\frac{1}{p}\frac{\partial}{\partial\xi}\left(\sqrt{1-\xi^{2}}S_{\xi}\right)=0,\label{eq:kineticexp}
\end{equation}
where the guiding-center momentum space flux components are
\begin{align}
\begin{split}
S_{p} & =-D_{pp,\textrm{FP}}\frac{\partial f}{\partial p}+\left(K_{p,\textrm{FP}}+K_{p,\textrm{E}}+K_{p,\textrm{ALD}}\right)f,\\
S_{\xi} & =\sqrt{1-\xi^{2}}D_{\xi\xi,\textrm{FP}}\frac{\partial f}{\partial\xi}+\left(K_{\xi,\textrm{E}}+K_{\xi,\textrm{ALD}}\right)f.
\label{eq:kineticfluxes}
\end{split}
\end{align}
The terms contributing to these fluxes are the convection and diffusion coefficients associated with
the Fokker-Planck collision operator (which are independent of $\xi$
for isotropic field particle distributions \cite{bra87,shk97})
\begin{alignat}{2}
D_{pp,\textrm{FP}} & = & A_{\textrm{FP}}\left(p\right),\label{eq:dpp}\\
K_{p,\textrm{FP}} & = & -F_{\textrm{FP}}\left(p\right),\\
D_{\xi\xi,\textrm{FP}} & = & \frac{B_{\textrm{FP}}\left(p\right)}{p},
\end{alignat}
the electric field acceleration
\begin{eqnarray}
K_{p,\textrm{E}} & = & \xi E_{\Vert},\\
K_{\xi,\textrm{E}} & = & -\sqrt{1-\xi^{2}}E_{\Vert},
\end{eqnarray}
and the synchrotron reaction force 
\begin{alignat}{2}
K_{p,\textrm{ALD}} & = & -\sigma_{r}\gamma p\left(1-\xi^{2}\right),\\
K_{\xi,\textrm{ALD}} & = & -\sigma_{r}\dfrac{p\xi\sqrt{1-\xi^{2}}}{\gamma}.\label{eq:kxi}
\end{alignat}

\noindent In Eqs.~(\ref{eq:kineticexp}-\ref{eq:kxi}), time is normalized to
the collision time for relativistic electrons, 
\begin{equation}
\tau_{c}=\frac{4\pi\varepsilon_{0}^{2}m^{2}c^{3}}{e^{4}n\ln\Lambda},
\end{equation}
momentum $p$ is given in units of $mc$, the parallel electric field
$E_{\Vert}$ is normalized to the critical field $E_{c}$, and $\sigma_{r}\equiv\tau_{c}/\tau_{r}$
measures the relative strength (compared to collisional forces) of the ALD radiation reaction force
\begin{equation}
\sigma_{r}=\frac{2}{3}\frac{1}{\ln\Lambda}\frac{\omega_{c}^{2}}{\omega_{p}^{2}},
\end{equation}
where $\omega_{p}$ is the electron plasma frequency defined by $\omega_{p}^{2}=e^{2}n/(\varepsilon_{0}m)$.
The collisional diffusion coefficients $A_{\textrm{FP}}\left(p\right)$
and $B_{\textrm{FP}}\left(p\right)$ are normalized to $(mc)^{2}/\tau_{c}$
while the friction coefficient $F_{\textrm{FP}}\left(p\right)$ is
normalized to $mc/\tau_{c}$. 

An explicit form of the momentum-space fluxes (\ref{eq:kineticfluxes})
is thus 
\begin{align}
\begin{split}
S_{p} & =-A_{\textrm{FP}}\left(p\right)\frac{\partial f}{\partial p}+\left[\xi E_{\Vert}-F_{\textrm{FP}}\left(p\right)-\sigma_{r}\gamma p\left(1-\xi^{2}\right)\right]f,\\
S_{\xi} & =\sqrt{1-\xi^{2}}\left(\frac{B_{\textrm{FP}}\left(p\right)}{p}\frac{\partial f}{\partial\xi}-E_{\Vert}-\sigma_{r}\gamma^{-1}p\xi\right)f.
\label{eq:kineticfluxesexp}
\end{split}
\end{align}
We assume cold and infinitely massive ions,
so that the normalized collision coefficients $A_{\textrm{FP}}$,
$F_{\textrm{FP}}$ and $B_{\textrm{FP}}$
only depend on $T_{e}$ and $Z_{\textrm{eff}}$ \cite{shk97}. Collisions with ions
only enter the pitch-angle scattering term $B_{\textrm{FP}}\left(p\right)=B_{\textrm{FP},e}\left(p\right)+Z_{\textrm{eff}}/(2v)$,
while $A_{\textrm{FP}}\left(p\right)=\beta^{2}F_{\textrm{FP}}\left(p\right)/v$,
where the normalized electron temperature is defined as $\beta^{2}\equiv k_BT_{e}/(mc^{2})$.
To summarize, the normalized equation (\ref{eq:kineticfluxesexp})
depends on the following independent parameters only: the parallel
electric field $E_{\Vert}$, the normalized ALD frequency $\sigma_{r}$,
the electron temperature $T_{e}$, and the effective charge $Z_{\textrm{eff}}$.

\section{Evolution of the electron distribution function\label{sec:evolution}}

\subsection{Force balance and runaway region}

Some preliminary insight into runaway electron dynamics can be extracted
from the force balance, $K_{p}\equiv K_{p,\textrm{FP}}+K_{p,\textrm{E}}+K_{p,\textrm{ALD}}=0$,
which can be expressed as 
\begin{equation}
\xi E_{\Vert}-F_{\textrm{FP}}\left(p\right)-\sigma_{r}\gamma p\left(1-\xi^{2}\right)=0.\label{eq:force-balance}
\end{equation}
In the high velocity limit (see Appendix \ref{sub:High-velocity-limit})
and in the absence of the ALD force, the force balance yields 
\begin{equation}
p^{2}=\frac{1}{\xi E_{\Vert}-1}.\label{eq:pcrit}
\end{equation}
For particles with purely parallel momentum, this condition determines
the critical momentum $p_{c}\equiv(E_{\Vert}-1)^{-1/2}$ above which
electrons are continuously accelerated. For particles with very large
momentum $p\gg1$, the condition (\ref{eq:pcrit}) provides an asymptotic
value $\xi_{c}=E_{\Vert}^{-1}$, such that $K_{p}>0$ for particles
with $\xi>\xi_{c}$.

In this paper, the runaway region is defined as the region where the
momentum force balance is positive, i.e. $K_{p}>0$. Electrons located
within this region are considered runaway electrons. In the absence
of the ALD force, this definition corresponds to the usual idea of a runaway
electron, as the probability for an electron to be continuously accelerated
if it enters the region where $K_{p,\textrm{FP}}+K_{p,\textrm{E}}>0$
is very high. The momentum space described by finite-difference Fokker-Planck
codes is a limited domain by nature with a high-energy boundary defined
by a maximum momentum $p_{\max}$. A proper description of the runaway
dynamics clearly requires $p_{\max}\gg p_{c}$. Then, we must distinguish between electrons located in the $K_{p,\textrm{FP}}+K_{p,\textrm{E}}>0$
region within the code simulation domain $p<p_{\max}$, \textit{internal
runaways}, and electrons having left the simulation domain,
\textit{ external runaways}. The total runaway
population $n_{r}$ consists of both internal and external runaways,
the latter being counted in the simulation, albeit without following
their momentum space characteristics.

When the ALD force is included, and for relativistic electrons with
$p_{\Vert}\gg1$ and $p_{\Vert}\gg p_{\bot}$, the force balance (\ref{eq:force-balance})
is approximately given by
\begin{equation}
E_{\Vert}-1-\sigma_{r}p_{\bot}^{2}=0
\end{equation}
and yields a condition on the perpendicular momentum $(p_{\bot}\equiv p\sqrt{1-\xi^{2}})$
\begin{equation}
p_{\bot0}^{2}=\frac{E_{\Vert}-1}{\sigma_{r}}.\label{eq:plim}
\end{equation}
The momentum space in the far tail of the distribution is separated
into the runaway region $p_{\bot}<p_{\bot0}$ where the electric force
dominates over the ALD force and the collisional drag such that the net
force is positive, and a region $p_{\bot}>p_{\bot0}$ where the ALD
force and collisional drag dominate the electric force such that the
net force is negative. 

As the calculations in the next sections will show, in the presence
of an ALD force, the probability for electrons with $K_{p}>0$ to escape 
the runaway region at some point is high. Therefore, the concept of a runaway 
electron in this case is more an extension of the usual definition than a true 
characteristic. We will also see that electrons labelled as runaways can be entirely kept
within the simulation domain such that there are no external runaways.

\subsection{The Fokker-Planck code LUKE}

The Fokker-Plank equation is solved numerically by
the relativistic guiding-center Fokker-Planck code
LUKE. Equation~(\ref{eq:gc_kinetic}) is discretized
in momentum space $(p,\xi)$ using a 2-D finite-difference scheme with
non-uniform grids and a 9-point differentiation procedure.  A total of
1200 points are used for the $p$ grid, with 140 grid points describing
the $0<p<3$ region with a constant grid step, and 1060 grid points
describing the $3<p<p_{\max}=200$ region using increasing grid steps
with cubic dependence. A total of 166 points are used for the $\xi$
grid, with a decreasing step size towards $\xi=\pm1$ for increased resolution near the $p_{\Vert}$ axis. The code LUKE has been
benchmarked for the usual runaway problem \cite{dec04a}.
A benchmark including the ALD radiation reaction force is
conducted against the Fokker-Planck solver CODE and presented in
Sec.~\ref{sub:Benchmark}.

The linearization of the electron-electron collision operator implies
that the calculation is valid only as long as the electron
distribution is not too distorted from the original Maxwellian, which
in practice implies that the runaway fraction $n_{r}/n$ does not
exceed a few percent.

\subsection{Time evolution of the distribution function} \label{sub:time_evo_of_dist}

\begin{figure*}
  \centering
  \subfigure[]{
    \centering
    \includegraphics[width=0.45\textwidth]{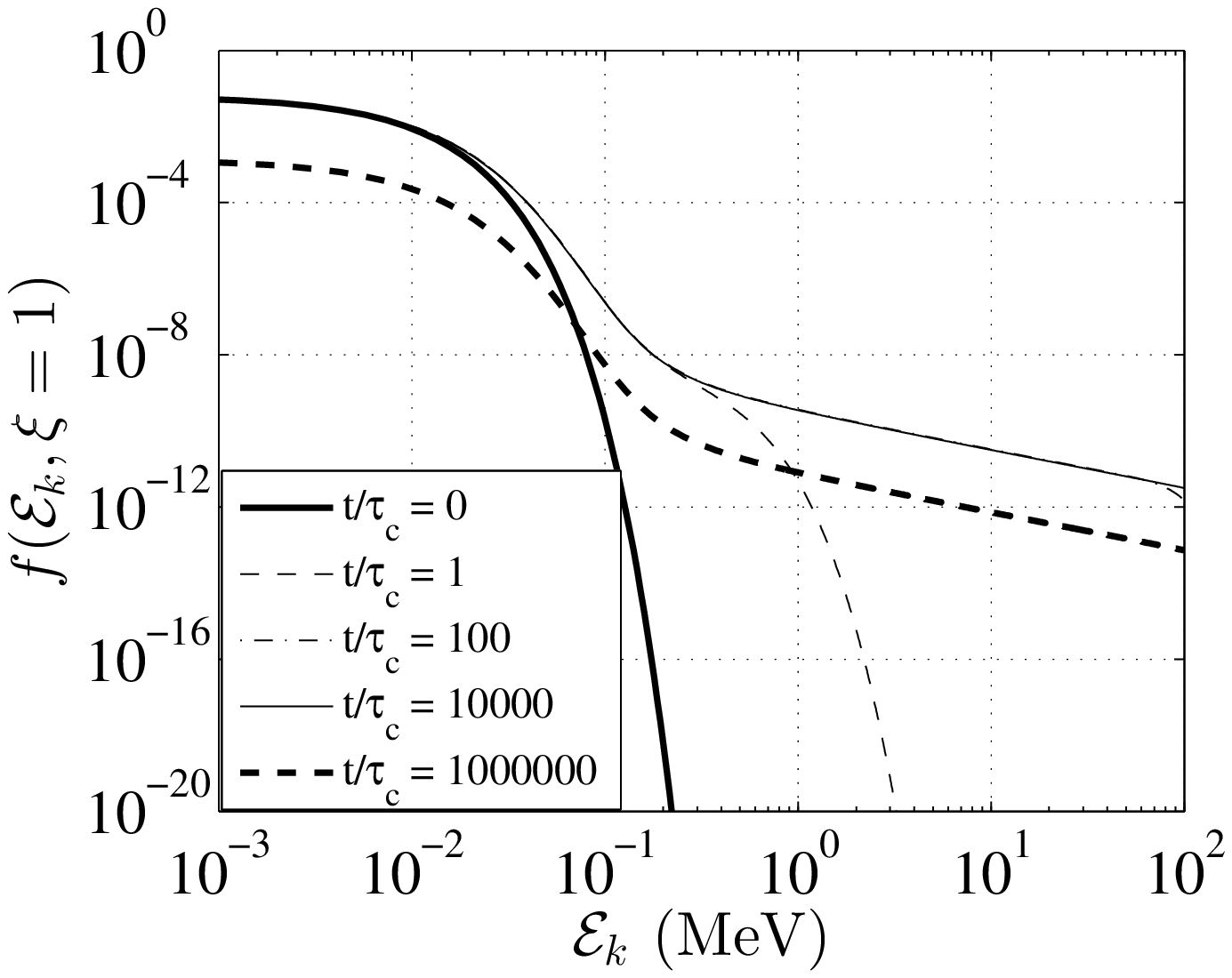}
    \label{fig:fevol-a}
  }
  \subfigure[]{
    \centering
    \includegraphics[width=0.45\textwidth]{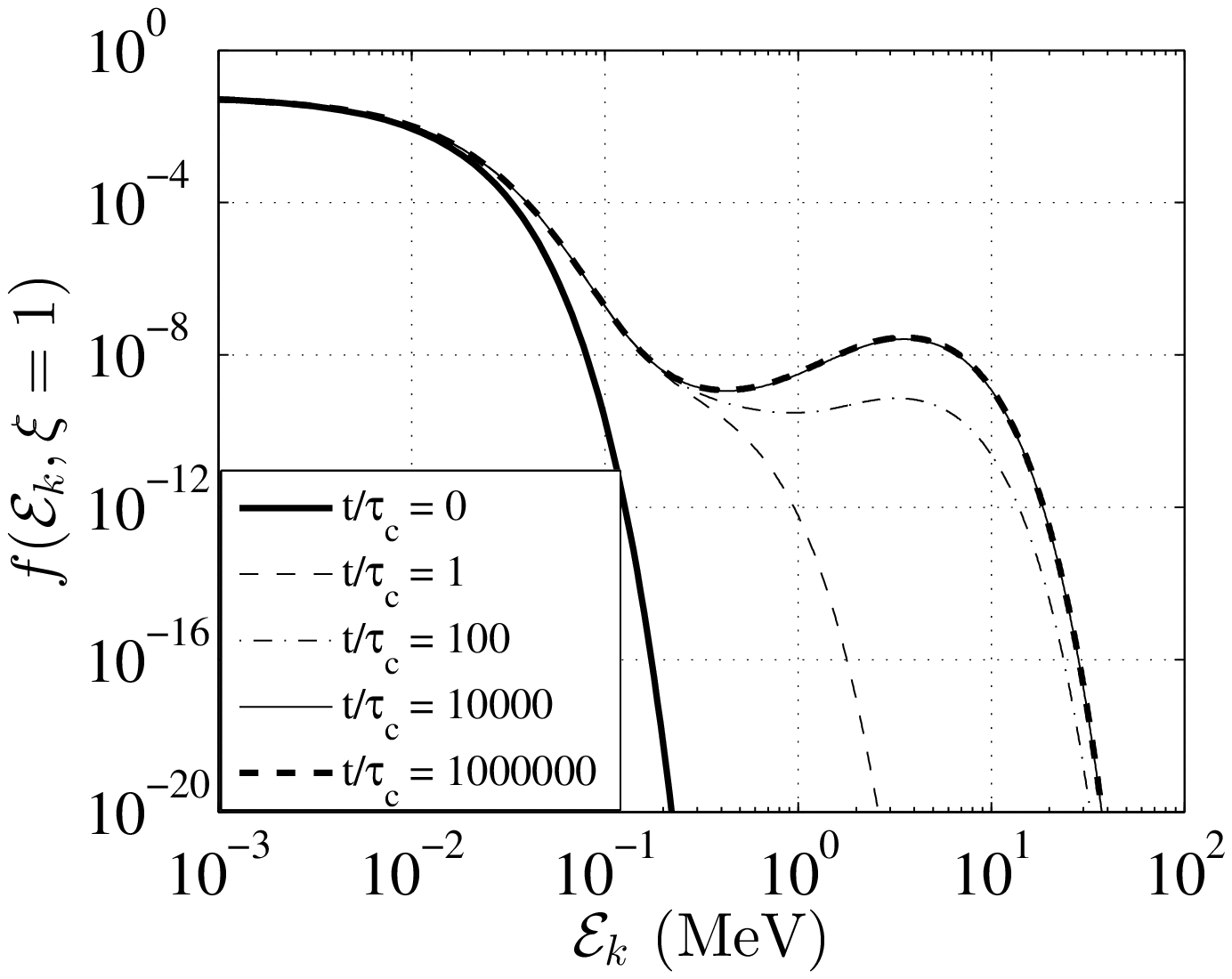}
    \label{fig:fevol-b}
  }
  \subfigure[]{
    \centering
    \includegraphics[width=0.45\textwidth]{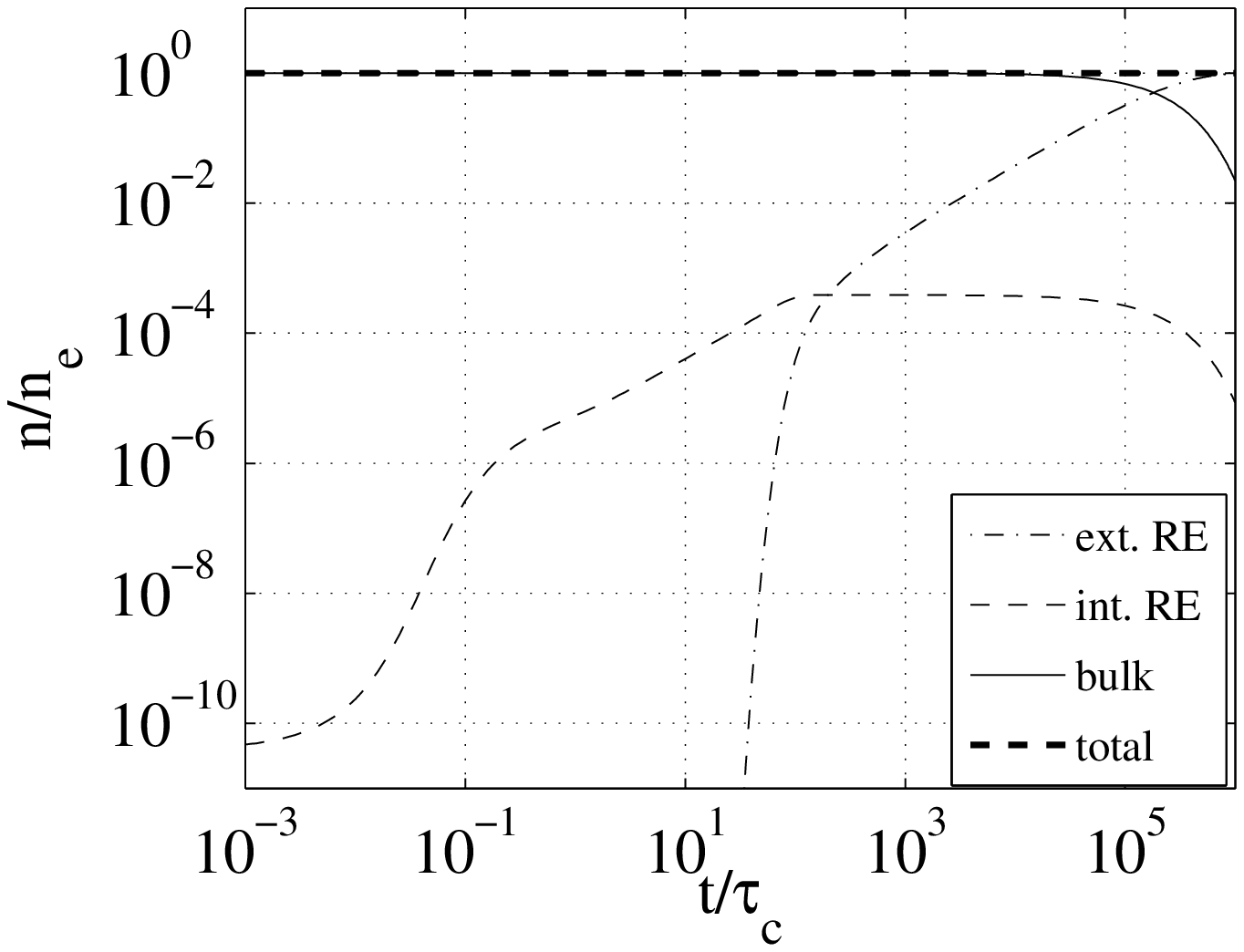}
    \label{fig:fevol-c}
  }
  \subfigure[]{
    \centering
    \includegraphics[width=0.45\textwidth]{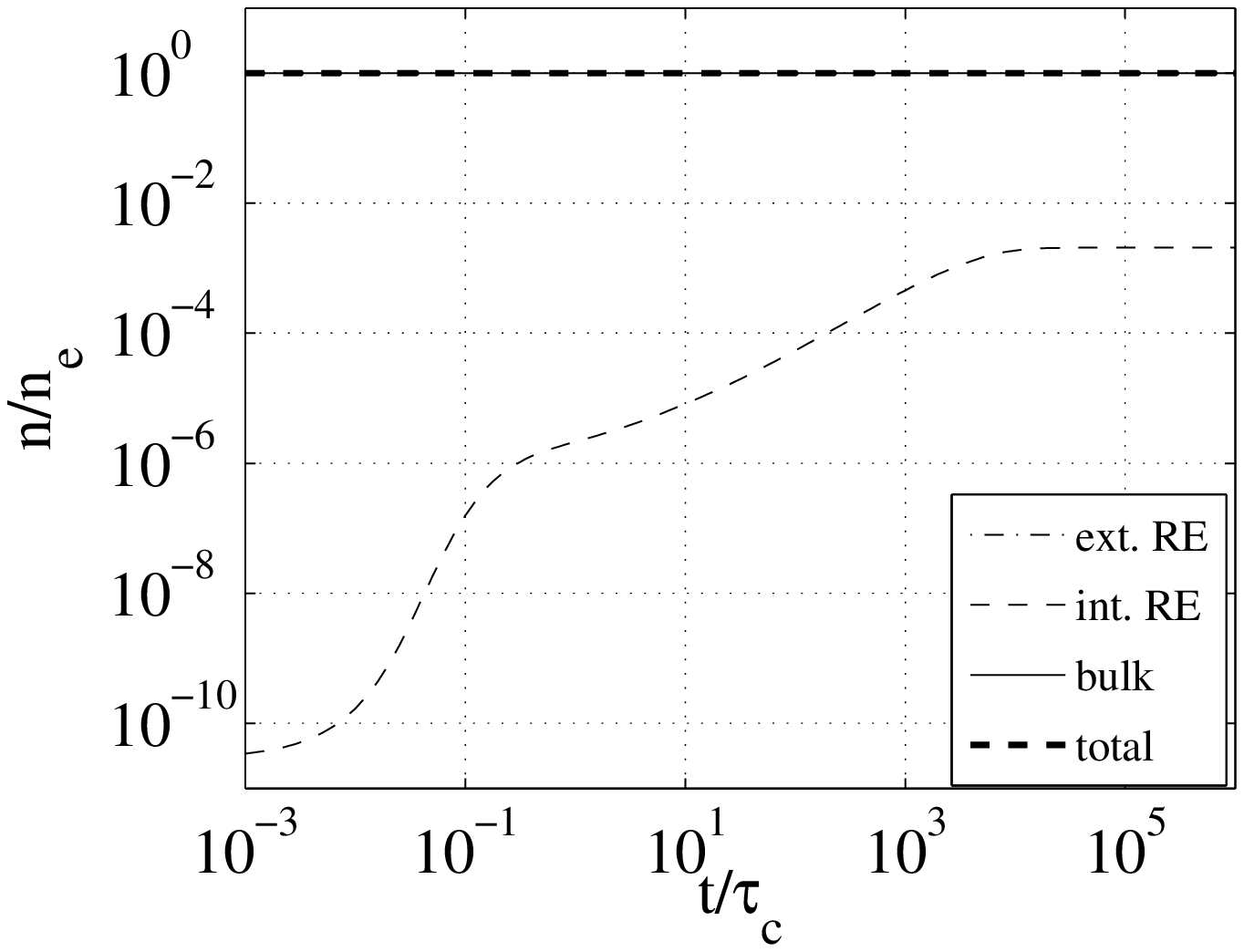}
    \label{fig:fevol-d}
  }
  \subfigure[]{
    \centering
    \includegraphics[width=0.45\textwidth]{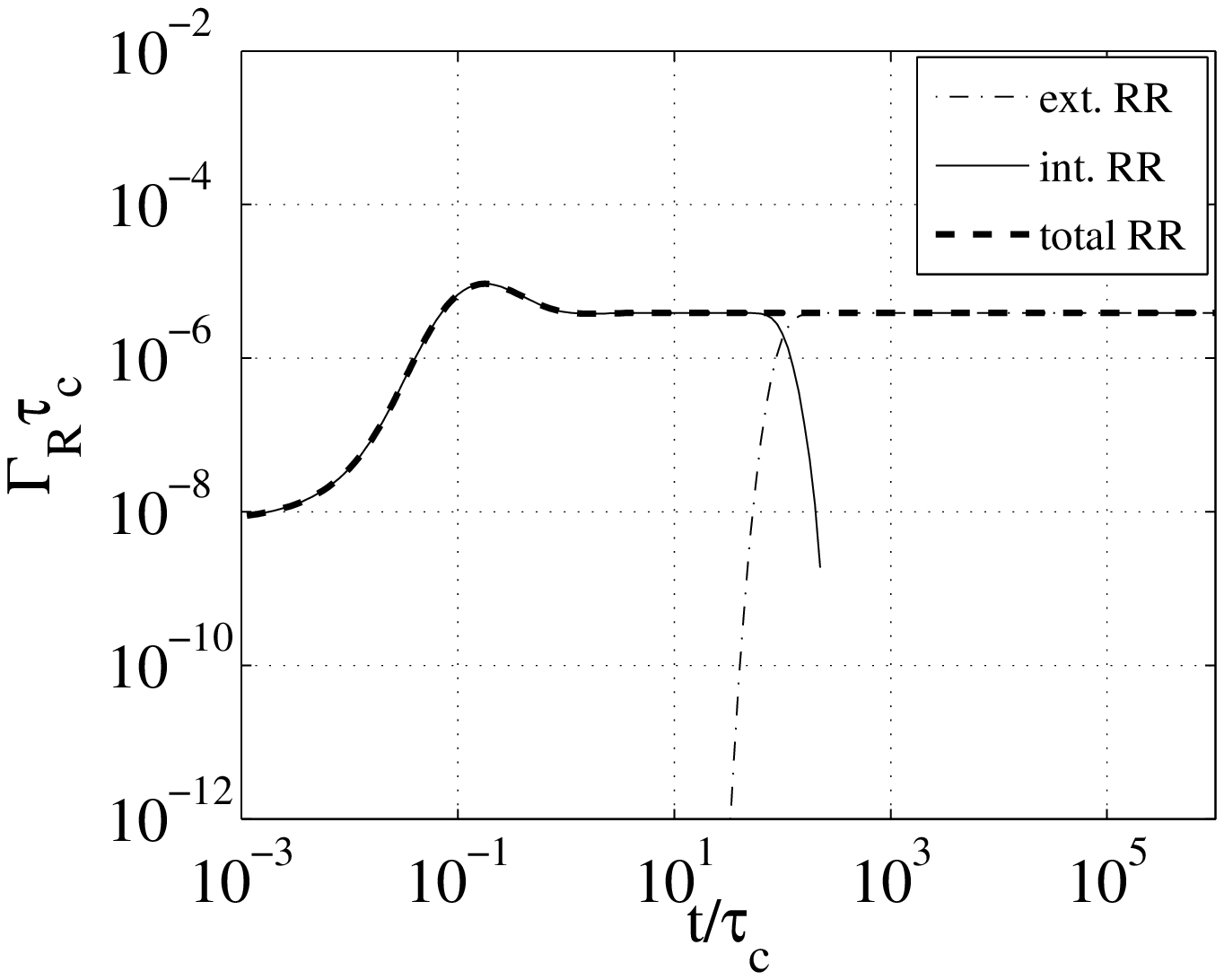}
    \label{fig:fevol-e}
  }
  \subfigure[]{
    \centering
    \includegraphics[width=0.45\textwidth]{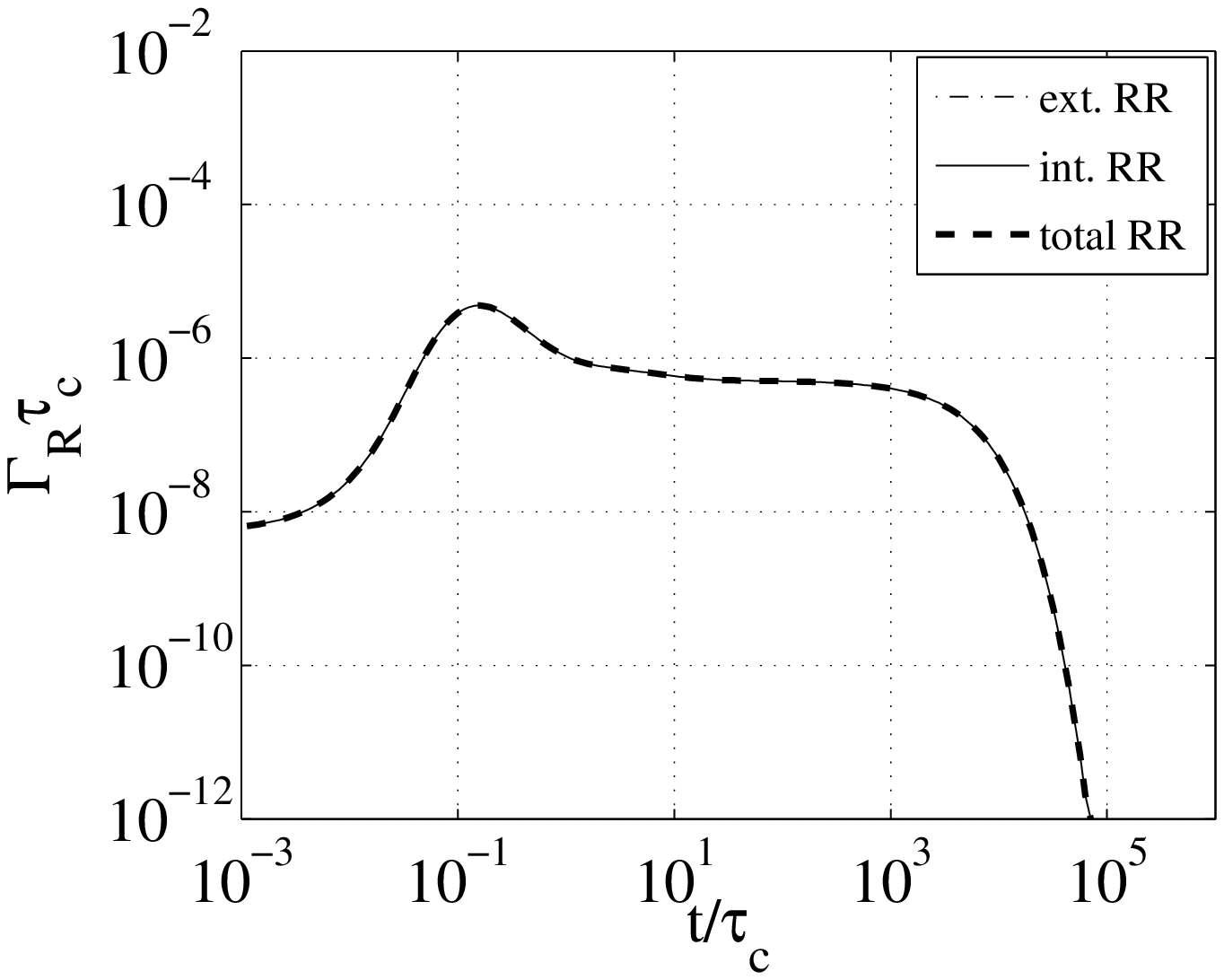}
    \label{fig:fevol-f}
  }

  \caption{Graphs (a) and (b) represent the evolution of the electron distribution
    function in the parallel direction ($\xi=1$) as a function of the
    electron kinetic energy $\mathcal{E}_{k}$; graphs (c) and (d) show the fraction of runaway
    electrons (RE) inside and outside the simulation domain,
    and graphs (e) and (f) the corresponding runaway rates. The parameters are
    $E_{\Vert}=3$ , $Z_{\textrm{eff}}=1$, $\beta=0.1$, and
    $\sigma_{r}=0.6$. The ALD force contribution is neglected in graphs
    (a,c,e), whereas it is included in graphs (b,d,f).}
  \label{fig:fevol}
  
\end{figure*}

The electron distribution evolves from an initial relativistic
Maxwellian distribution, which is also the steady-state solution of
Eq.~(\ref{eq:kineticexp}) for $E_{\Vert}=0$ and $\sigma_{r}=0$. A constant
electric field $E_{\Vert}=3$ is applied, the effective charge is
$Z_{\textrm{eff}}=1$, the temperature is $\beta=0.1$ ($T_e=5.11$ keV), and the
ratio $B^{2}/n$ is adjusted such that $\sigma_{r}=0.6$, which
corresponds to typical low-density conditions in tokamak plasmas
(i.e. $n=10^{19}$ m$^{-3}$ and $B=4$~T in the Tore-Supra tokamak). The
evolution of the electron distribution function in the parallel
direction ($\xi=1$) is shown in Fig.~\ref{fig:fevol}.

In the absence of the ALD force, a runaway tail progressively extends to
the edge of the simulation box (Fig.~\ref{fig:fevol-a}). The electron
distribution does not converge to a steady-state. The runaway rate
reaches an asymptotic value (Fig.~\ref{fig:fevol-e}), and the fraction
of runaway electrons increases continuously (Fig.~\ref{fig:fevol-c}).
At first, the runaway rate is related to an increase in the internal
runaway electron population. Once the runaway tail reaches the edge of
the simulation domain (for $t\sim200$), the runaway rate is related to
the population leaving the simulation box and becoming external
runaways. Note that, strictly speaking, the linearized collision
operator is no longer valid for $t/\tau_{c}>10^{5}$ as the fraction of
runaway electrons becomes of order unity. Nevertheless, the evolution is continued to illustrate the
absence of a steady-state solution. In summary, the distribution
function (Fig.~\ref{fig:fevol-a}) evolves towards an asymptotic
solution with the bulk population depleting at a constant rate.

In the presence of the ALD force, however, we find that the
distribution evolves towards a steady-state solution
(Fig.~\ref{fig:fevol-b}) as the runaway rate vanishes
(Fig.~\ref{fig:fevol-f}). No electron leaves the simulation box, and
the population of internal runaways reaches an asymptotic value
$n_{r}/n=0.002$ (Fig.~\ref{fig:fevol-d}).  In addition, we can observe
the formation of a region with positive gradient in parallel momentum,
which appears as a high-energy bump in the tail of the distribution
function (Fig.~\ref{fig:fevol-b}).  Properties of the time-asymptotic
electron distribution function are examined in the next section.

\section{Steady-state solution and bump formation \label{sec:Properties}}

\subsection{Steady-state solution\label{sub:Steady-state-solution}}

\begin{figure}
  \centering
  \subfigure[]{
    \centering
    \includegraphics[width=0.45\textwidth]{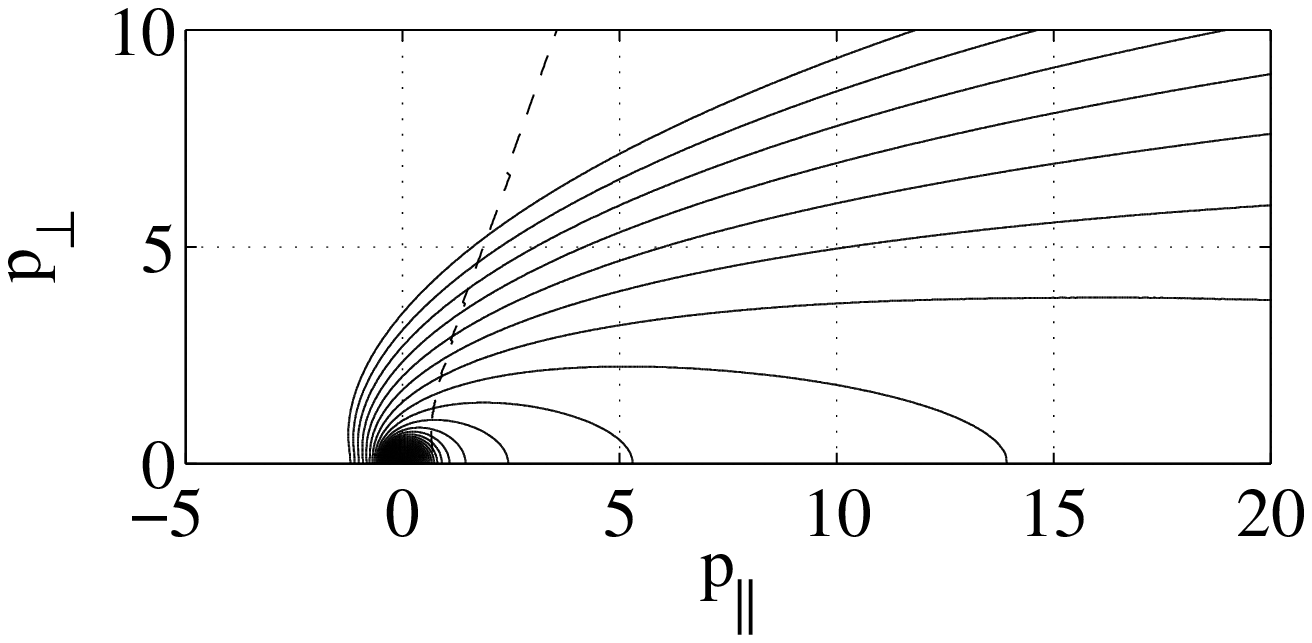}
    \label{fig:f2D-a}
  }
  \subfigure[]{
    \centering
    \includegraphics[width=0.45\textwidth]{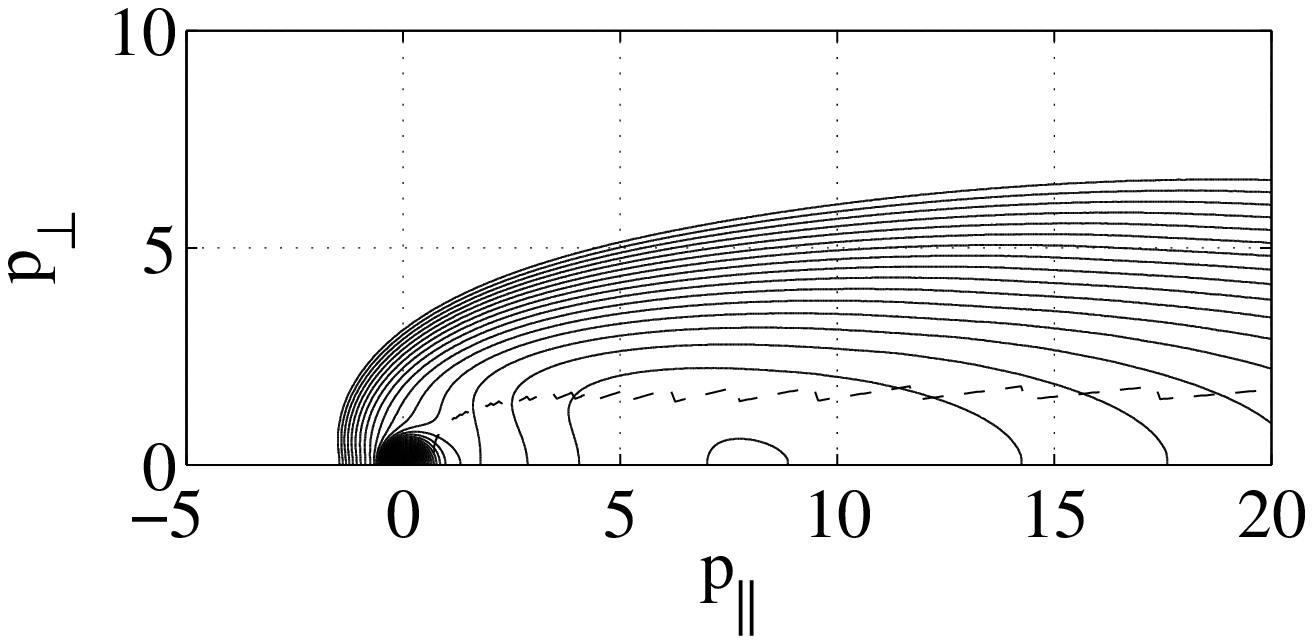}
    \label{fig:f2D-b}
  }
  \subfigure[]{
    \centering
    \includegraphics[width=0.45\textwidth]{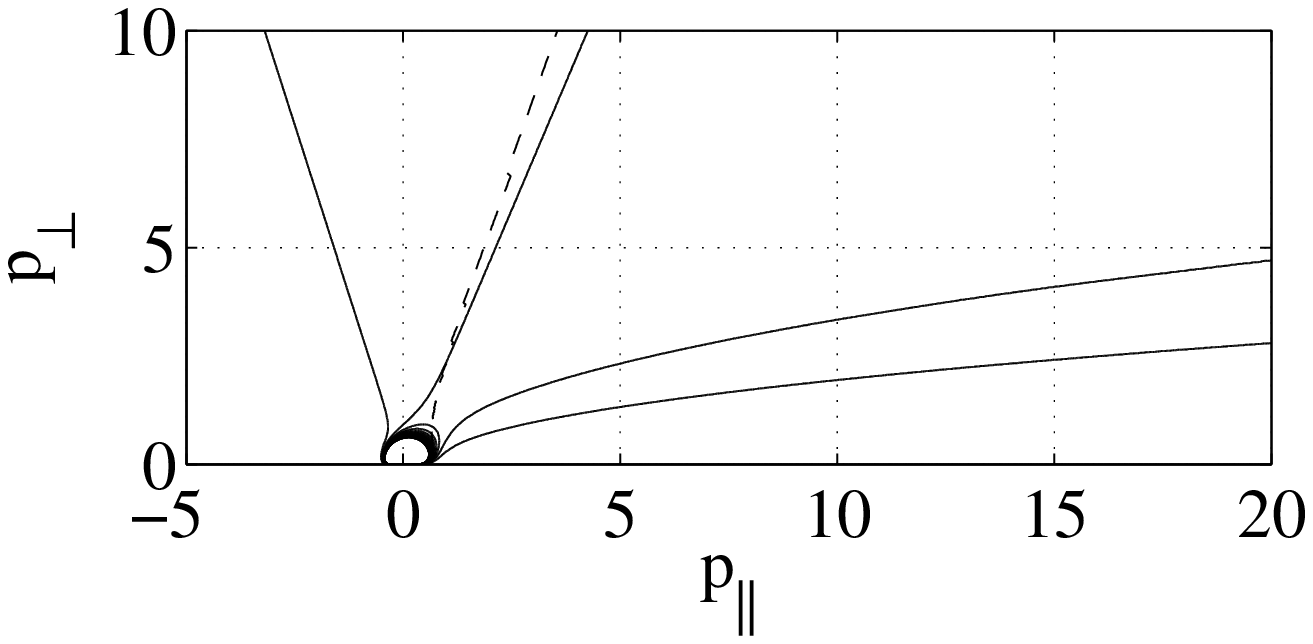}
    \label{fig:f2D-c}
  }
  \subfigure[]{
    \centering
    \includegraphics[width=0.45\textwidth]{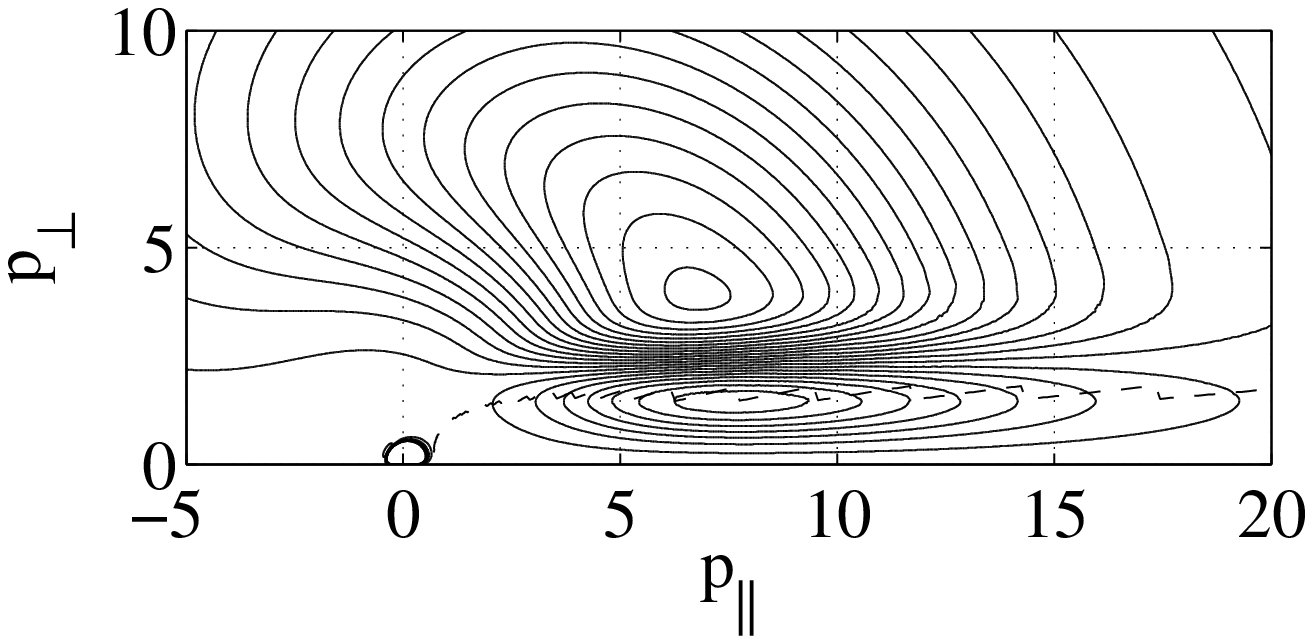}
    \label{fig:f2D-d}
  }
  \caption{Contours of the electron distribution function in graphs (a) and (b)
    and stream function in graphs (c) and (d) in 2-D guiding-center momentum space
    $(p_{\Vert},p_{\bot})$ at time $t/\tau_{c}=10^{6}$. The parameters
    are $E_{\Vert}=3$ , $Z_{\textrm{eff}}=1$, $\beta=0.1$, and $\sigma_{r}=0.6$.
    The ALD force contribution is neglected in (a) and (c) whereas it
    is accounted for in (b) and (d).}
  \label{fig:f2D}
\end{figure}

If a steady-state solution exists, it satisfies the equation
$\boldsymbol{\nabla}_{p,\xi}\cdot\mathbf{S}_{p,\xi}=0$.  Given the
axi\-symmetry of the momentum space, the
divergence-free steady-state fluxes can be expressed as
\begin{equation}
\mathbf{S}_{p,\xi}=\boldsymbol{\nabla}_{p,\xi}\times\left[\frac{A(p,\xi)}{2\pi p\sqrt{1-\xi^{2}}}\hat{\boldsymbol{\varphi}}\right],
\end{equation}
where $A(p,\xi)$ is called the stream function. Since $\mathbf{S}_{p,\xi}\cdot\boldsymbol{\nabla}_{p,\xi}A=0$,
contours of $A(p,\xi)$ indicate the direction of the momentum-space
fluxes, or streamlines. The total flux of electrons between two contours
is given by the corresponding difference in the value of $A(p,\xi)$,
such that narrowing contours indicate regions of stronger flux. The
stream function thus provides a very informative graphical representation
of the steady-state fluxes in momentum space. While no steady-state
solution to the runaway problem exists in the absence of ALD force,
it is possible to artificially obtain a steady-state distribution
function by adding a source term at $p=0$, which compensates exactly
for the external runaway rate. Whereas adding cold electrons does
not change the runaway rate or the shape of the distribution function
in the tail, it enables us to interpret the stream function as a representation
of the steady-state fluxes. 

The 2D representation of the steady-state solution corresponding to
the simulation parameters from Section~\ref{sub:time_evo_of_dist} is presented
in Fig.~\ref{fig:f2D}. The distribution function is shown in Figs.~\ref{fig:f2D-a} and~\ref{fig:f2D-b} 
for the case without and with ALD force, respectively.
The dashed line delimits the runaway region where $K_{p}>0$. The corresponding
contours of the stream function $A(p,\xi)$ are drawn in Figs.~\ref{fig:f2D-c} and~\ref{fig:f2D-d}.

In the absence of the ALD force, the runaway population peaks near the
$p_{\Vert}$ axis but extends quite far in the perpendicular direction
into the runaway region, as seen if Fig.~\ref{fig:f2D-a}. The open
streamlines represented by the contours of the stream function in
Fig.~\ref{fig:f2D-c} show that electrons located in the runaway region
are indefinitely accelerated and eventually escape the simulation
domain. 

In the presence of the ALD force, the runaway region consists of a
narrow band along the $p_{\Vert}$ axis delimited by $p_{\bot}<p_{\bot0}$
at large energies. As seen by the contours of the stream function in Fig
\ref{fig:f2D-d}, the steady-state electron flux is directed towards
higher energies for $p_{\bot}<p_{\bot0}$ and towards lower energies
for $p_{\bot}>p_{\bot0}$. The electron tail is thus confined in the
region close to $p_{\bot}=0$, as seen in Fig.~\ref{fig:f2D-b}, which
creates a strong gradient in $p_{\bot}$. As seen in the next section,
this gradient results in strong pitch-angle scattering,
which contributes to limiting the energy of runaway electrons and
gives rise to a bump in the distribution under certain conditions. This
bump is centered on the $p_{\Vert}$ axis. As the minimum between
the bulk and the bump population naturally lies in the vicinity of
the critical field, the bump population almost coincides with the
runaway region, such that the electron population can be formally
separated into a bulk and a runaway ``beam''.

\subsection{Perpendicular force balance}

The spherical representation $(p,\xi)$ is the natural coordinate
system for describing collisions. Thus, it is used for the numerical discretization
of the kinetic equation in the Fokker-Planck code LUKE.
As seen in Fig.~\ref{fig:f2D-b}, however, the tail of the distribution
function is determined by the runaway region $p_{\bot}<p_{\bot0}$
and the natural coordinate system is rather the cylindrical representation
$(p_{\bot},p_{\Vert})$ with $p_{\bot}=p\sqrt{1-\xi^{2}}$ and $p_{\Vert}=p\xi$. 

The transformation to the $(p_{\bot},p_{\Vert})$ is detailed in Section
\ref{sub:Cylindrical-representation}. Focusing on the tail region
of momentum space near the parallel axis (characterized by $p_{\bot}\ll p_{\Vert}$
and $p_{\Vert}\gg\beta$), the momentum-space fluxes entering (\ref{eq:kineticfluxescyl})
yield to leading order in $\beta$
\begin{align}
\begin{split}S_{\bot} & =-\frac{1+Z_{\textrm{eff}}}{2v}\frac{\partial f}{\partial p_{\bot}}-\frac{p_{\bot}}{p_{\Vert}}\left[\frac{1}{v^{2}}+\sigma_{r}v\left(1+p_{\bot}^{2}\right)\right]f,\\
S_{\Vert} & =\left[E_{\Vert}-\frac{1}{v^{2}}-\sigma_{r}vp_{\bot}^{2}\right]f,
\label{eq:kineticfluxescyltail}
\end{split}
\end{align}
where $v\simeq p_{\Vert}/\gamma\simeq p_{\Vert}/\sqrt{1+p_{\Vert}^{2}}$ 
is the velocity normalized to the speed of light.

At high energy, $v\simeq1$ and (\ref{eq:kineticfluxescyltail}) becomes
approximately
\begin{align}
\begin{split}S_{\bot} & =-\frac{1+Z_{\textrm{eff}}}{2}\frac{\partial f}{\partial p_{\bot}}-\frac{p_{\bot}}{p_{\Vert}}\left[1+\sigma_{r}\left(1+p_{\bot}^{2}\right)\right]f,\\
S_{\Vert} & =\sigma_{r}\left[p_{\bot0}^{2}-p_{\bot}^{2}\right]f.
\label{eq:kineticfluxescyltail2}
\end{split}
\end{align}
The parallel flux of electrons $S_{\Vert}$ is positive for particles
with $p_{\bot}<p_{\bot0}$ and negative for particles with $p_{\bot}>p_{\bot0}$.
The resulting strong perpendicular gradient enhances pitch-angle scattering,
which creates a positive flux in the $p_{\bot}$ direction at high
energy, which is illustrated by the streamlines in Fig.~\ref{fig:f2D-d}.
This flux limits the extension of the electron distribution to higher
energies, as seen in Fig.~\ref{fig:f2D-b}. The existence of a bump
in the distribution is also driven by the perpendicular dynamics.
From the expression (\ref{eq:kineticfluxescyltail2}) for $S_{\bot}$,
we see that the convective component decreases with $p_{\Vert}$ while
pitch-angle scattering is independent of $p_{\Vert}$ for a given
perpendicular gradient $\partial f/\partial p_{\bot}$. Therefore,
on average, electrons in the tail are pushed into the runaway region
at lower $p_{\Vert}$, while they are scattered away at higher $p_{\Vert}$.
This dynamics is clearly seen in the stream function plot \ref{fig:f2D-d}.
Naturally, the bump appears at the balance point where $S_{\bot}\simeq0$
and electrons accumulate as a result of the perpendicular dynamics. Given
the cylindrical symmetry, and since the perpendicular gradient is created
by the parallel force balance, we may assume a parabolic dependence
scaled by $p_{\bot0}$ around the bump location
\begin{equation}
\frac{1}{f}\frac{\partial f}{\partial p_{\bot}}\propto-\frac{p_{\bot}}{p_{\bot0}^{2}}
\end{equation}
such that from (\ref{eq:kineticfluxescyltail2}) and for $p_{\bot}\rightarrow0$
we obtain an estimate for the parametric dependence of the position
of the bump
\begin{equation}
p_{\Vert b}\propto\frac{2}{1+Z_{\textrm{eff}}}\frac{1+\sigma_{r}}{\sigma_{r}}\left(E_{\Vert}-1\right),\label{eq:bumploc}
\end{equation}
which is in agreement with the expression obtained from approximately
solving the kinetic equation analytically \cite{hir15}. It is quite
intuitive to expect that the bump energy increases with the electric
field amplitude, whereas it decreases with the amplitude of the ALD
force and the effective charge. However, the underlying processes are rather complex
and involves both the parallel and perpendicular dynamics. 
The parametric dependence of the bump location predicted by Equation (\ref{eq:bumploc}) 
will be compared to numerical calculations in Section \ref{sec:Parametric}.

\subsection{Validity of the uniform plasma approximation}

In tokamak plasmas, particles with purely parallel velocity are subject
to an ALD force due to the toroidal and poloidal periodic motions.
For a safety factor $q\approx1$ and electrons with $p_{\Vert}\gg1$, 
the contribution from the field line 
curvature to the ALD force is derived in Appendix \ref{sub:parallel} 
and is expressed as (\ref{eq:kfinal})
\begin{equation}
K_{R}=-\sigma_{r}\left(\frac{\rho_{0}}{R}\right)^{2}p_{\Vert}^{4},
\end{equation}
where $\rho_{0}=mc/(eB)$ is the Larmor radius of relativistic electrons
and $R$ is the major radius. The momentum $p_{R}$ for which the
toroidal ALD and drag forces compensate the electric force is thus given
by $E_{\Vert}-1-\sigma_{r}(\rho_{0}/R)^{2}p_{R}^{4}=0$, which yields
\begin{equation}
p_{R}=\left(\frac{E_{\Vert}-1}{\sigma_{r}}\right)^{1/4}\left(\frac{R}{\rho_{0}}\right)^{1/2}.
\end{equation}
For the Tore-Supra example shown in Sec.~\ref{sec:evolution},
$\rho_{0}/R=1.5\times10^{-4}$ and we find $p_{R}=110$, which corresponds
to $55$ MeV electrons. We see in Fig.~\ref{fig:fevol-b} that the
combination of uniform plasma ALD force and pitch-angle scattering
limits the distribution to energies much below $55$ MeV. Toroidal
effects could thus be neglected in this case.

\subsection{Benchmark of the solution from the LUKE and CODE codes\label{sub:Benchmark}}

\begin{figure}
  \centering
  \subfigure[]{
    \centering
    \includegraphics[width=0.45\textwidth]{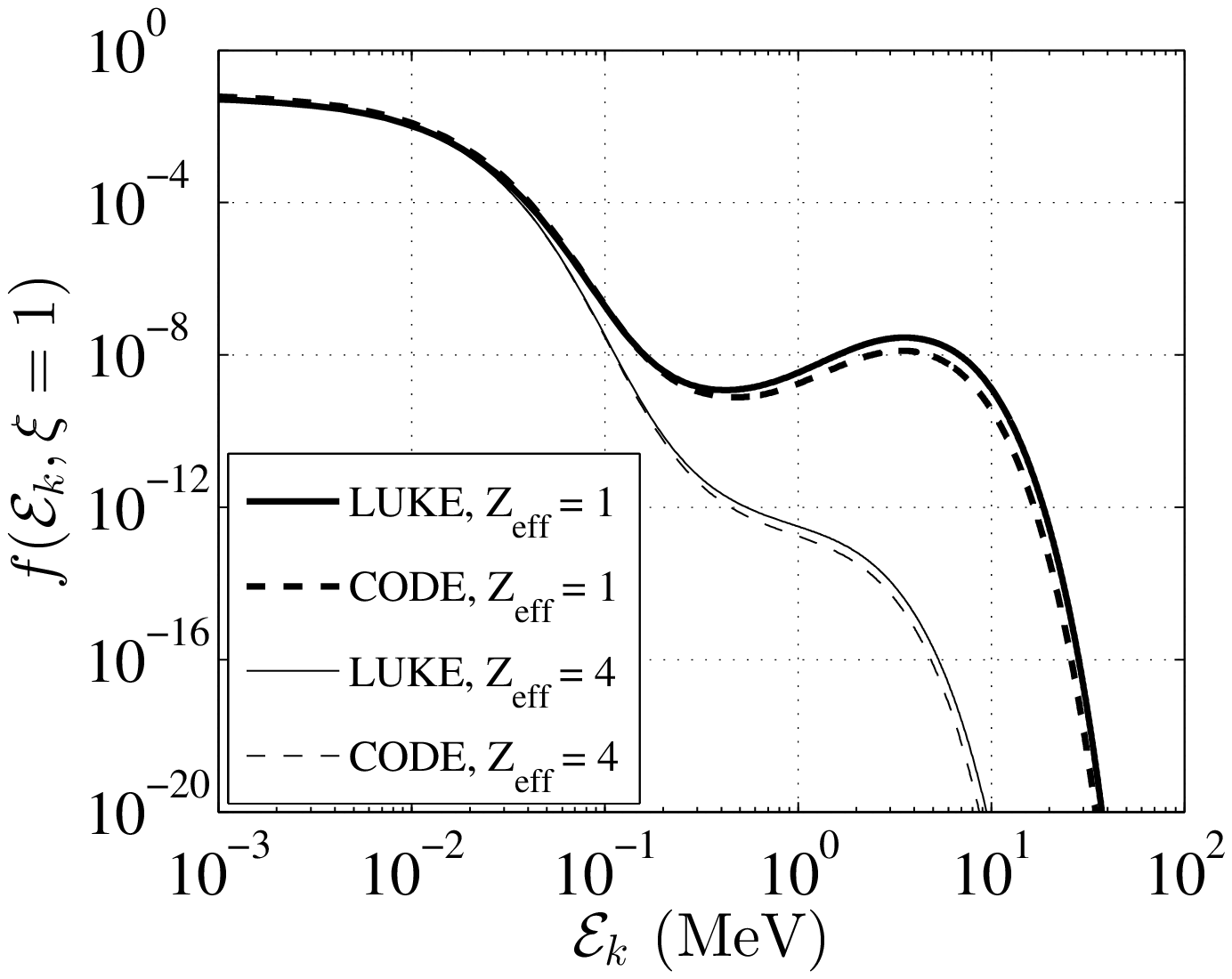}
    \label{fig:luke-code-a}
  }
  \subfigure[]{
    \centering
    \includegraphics[width=0.45\textwidth]{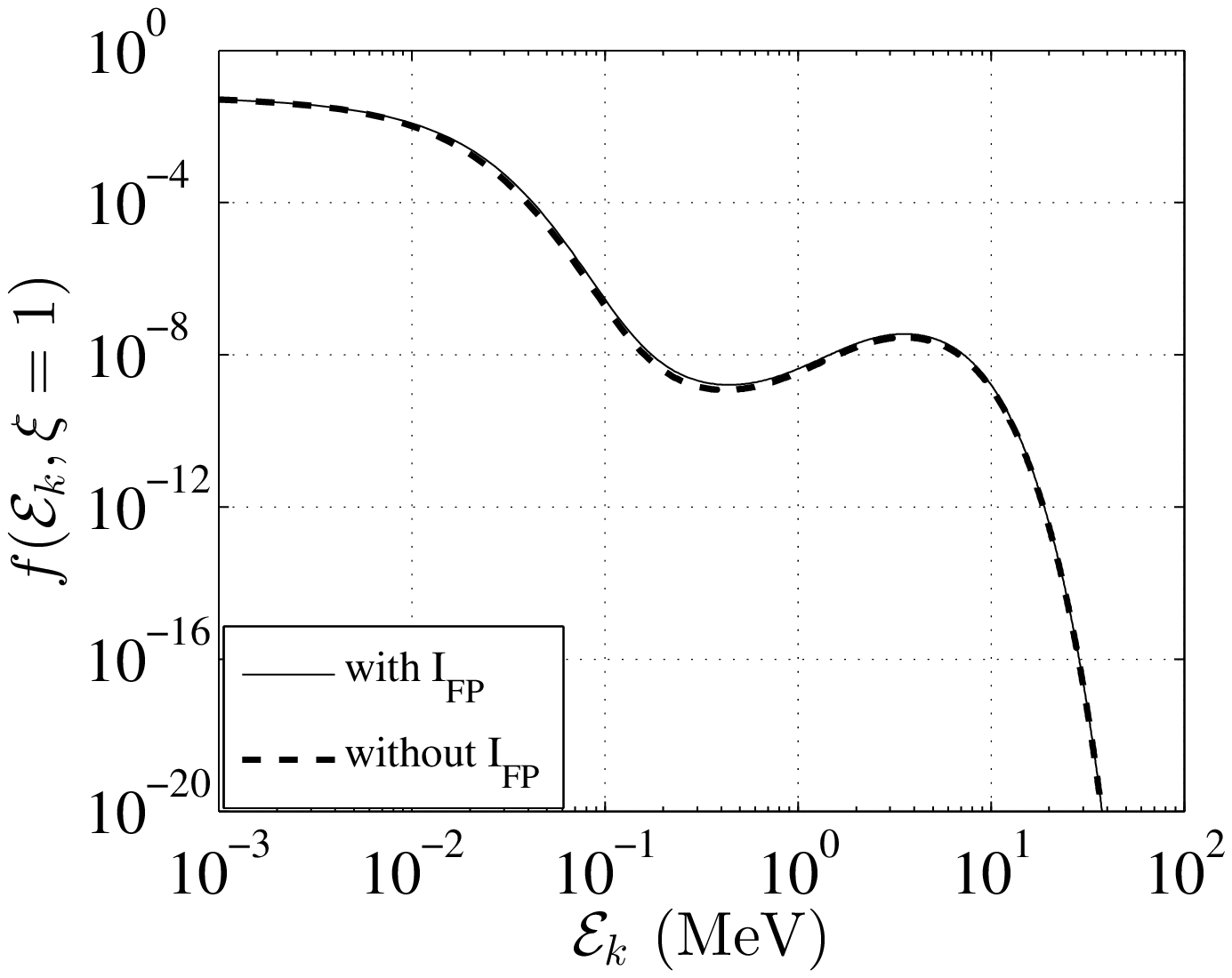}
    \label{fig:luke-code-b}
  }
  \caption{Electron distribution function in the parallel
    direction ($\xi=1$) as a function of the electron kinetic energy $\mathcal{E}_{k}$.
    Graph (a) compares results from the code LUKE and CODE for $Z_{\textrm{eff}}=1$
    and $Z_{\textrm{eff}}=4$, respectively. Graph (b) compares LUKE calculations
    in which the integral part of the collision operator $I_{\textrm{FP}}$
    is included in the Fokker-Planck equation (\ref{eq:gc_kinetic}) to the case where it is not.
    Fixed relevant parameters are $E_{\Vert}=3$, $\beta=0.1$, and $\sigma_{r}=0.6$. }
  \label{fig:luke-code}
\end{figure}

The simulations presented in this paper were obtained using the code
LUKE. While the code is extensively benchmarked for the usual runaway
problem \cite{dec04a}, LUKE simulations including the ALD reaction force are presented
for the first time in this paper. In order to benchmark the numerical
simulations, calculations from LUKE are compared to those from the
solver CODE, which solves the same Fokker-Planck equation
(\ref{eq:kineticexp}) but uses a spectral representation of 
the pitch-angle dependence \cite{lan14}.  
The corresponding steady-state distribution
functions are shown in Fig.~\ref{fig:luke-code-a} for the parameters
used in Sec.~\ref{sub:time_evo_of_dist}, and two different values for the
effective charge, $Z_{\textrm{eff}}=1$ and $Z_{\textrm{eff}}=4$.
Results from the two codes are in excellent agreement.
In particular, both codes show the appearance of a bump at the same energy for
$Z_{\textrm{eff}}=1$, while they show no bump formation for
$Z_{\textrm{eff}}=4$.

In addition, the integral part of the collision operator $I_{\textrm{FP}}$
can be included in the code LUKE. Comparing the cases with and without
$I_{\textrm{FP}}$, we find that the distribution functions are very
similar, as seen in Fig.~\ref{fig:luke-code-b}. This is not surprising
as $I_{\textrm{FP}}$ mainly affects the bulk population, such that
it is appropriate to ignore it in the context of the present paper.
However, as $I_{\textrm{FP}}$ ensures momentum conservation in the
electron-electron collision operator, it must be included for accurate
driven current calculations. Indeed, the current density associated
with the distributions shown in Fig.~\ref{fig:luke-code-b} is $J/(ecn)=0.015$
without $I_{\textrm{FP}}$ while it is $J/(ecn)=0.027$ when $I_{\textrm{FP}}$
is included. The difference arises from a shift of the electron bulk in the
parallel direction, which is hardly visible in Fig.~\ref{fig:luke-code-b}; however,
the resulting asymmetry has a strong effect on the corresponding current.

\section{Parametric dependences of the electron distribution\label{sec:Parametric}}

The relevant physical parameters for the runaway electron problem described in this paper
are the electric field amplitude, the magnitude of the ALD radiation reaction force, the 
effective charge, and the electron temperature. In this section, the bump formation is characterized 
as a function of these parameters. The distribution function is evolved until it reaches
a steady-state solution. 

\begin{figure}
  \centering
  \subfigure[]{
    \centering
    \includegraphics[width=0.45\textwidth]{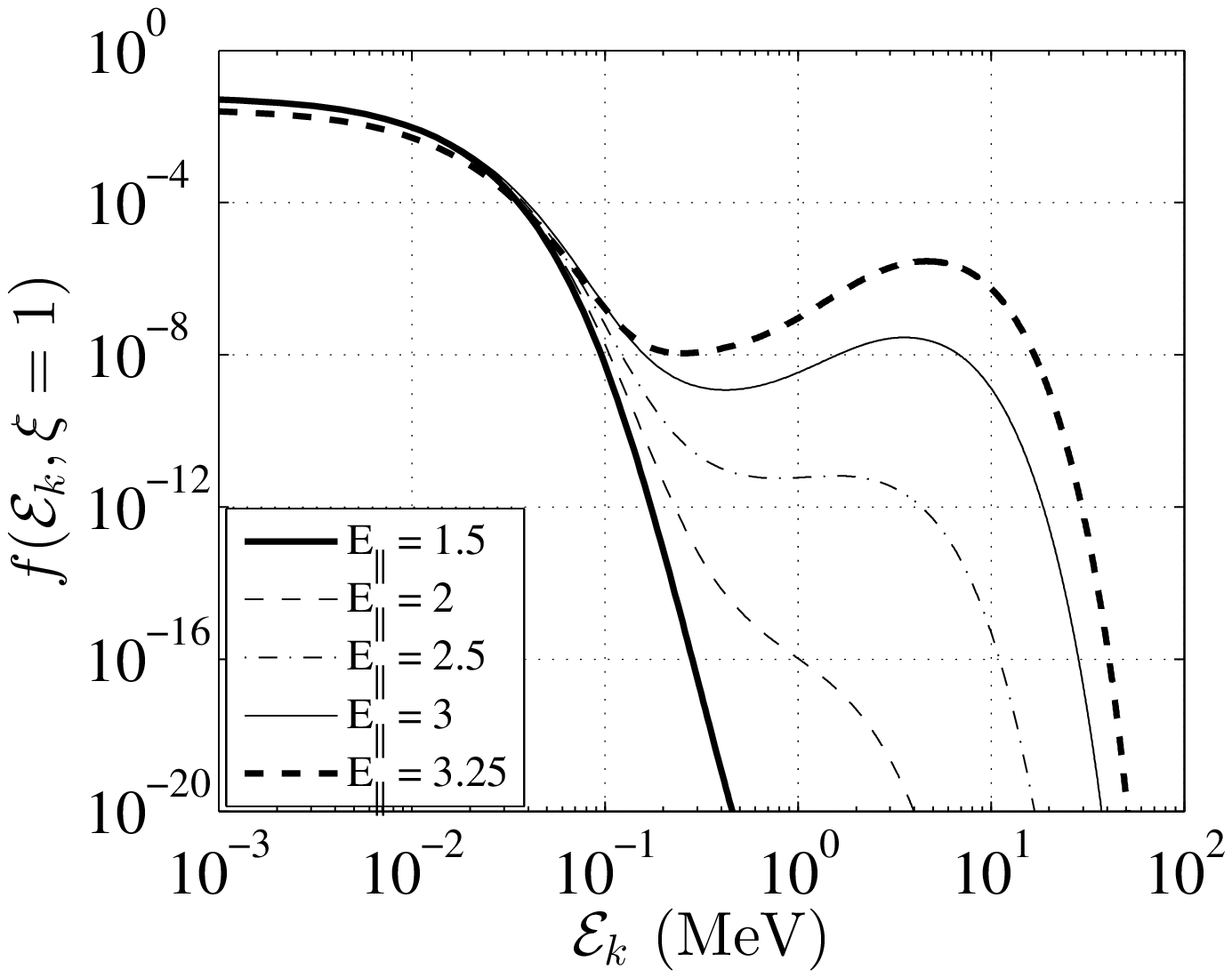}
    \label{fig:param_E-a}
  }
  \subfigure[]{
    \centering
    \includegraphics[width=0.45\textwidth]{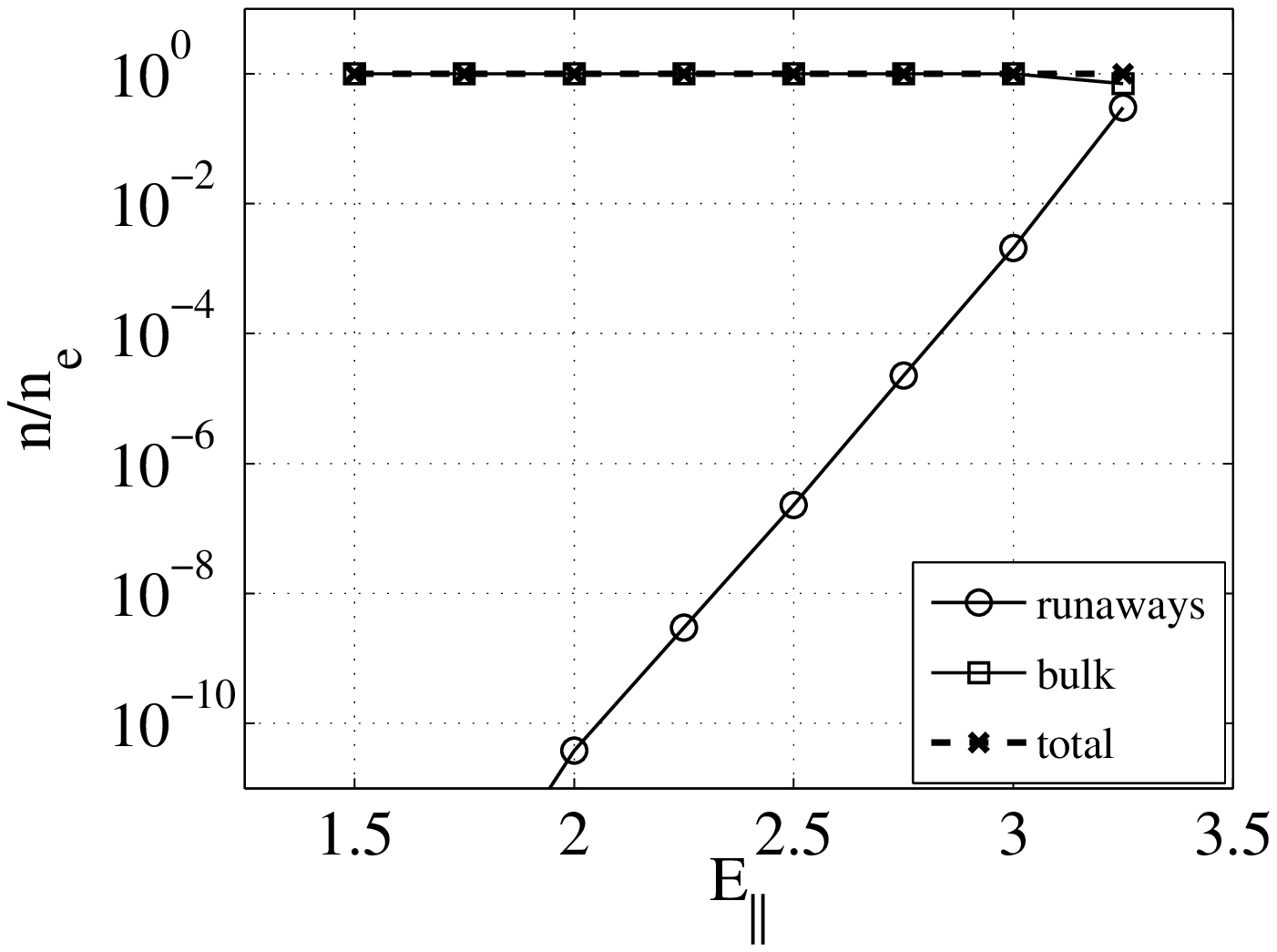}
    \label{fig:param_E-b}
  }
  \caption{(a) Electron distribution function in the parallel
    direction ($\xi=1$) as function of the electron kinetic energy $\mathcal{E}_{k}$, 
    and (b) fraction of electrons in the runaway region. The simulation
    results are plotted for various values of the parallel electric field.
    Fixed relevant parameters are $Z_{\textrm{eff}}=1$, $\beta=0.1$,
    and $\sigma_{r}=0.6$. }
  \label{fig:param_E}
\end{figure}

In a first set of calculations, the electric field is varied while
keeping the other relevant parameters fixed, with $Z_{\textrm{eff}}=1$,
$\beta=0.1$, and $\sigma_{r}=0.6$. The results are shown in Fig.
\ref{fig:param_E}. We observe that the electric field must reach
a certain threshold for the bump to appear in the tail of the distribution
function. Above this threshold, the energy corresponding to the bump
location increases with $E_{\Vert}$, in accordance with the estimate
(\ref{eq:bumploc}). In addition we observe that the number of runaway
electrons, i.e. the number of electrons with a positive parallel force
balance, increases with the amplitude of the electric field. Note
that the calculation was restricted to $E_{\Vert}<3.5$, as the linearization
of the collision operator fails above this limit since the runaway
population becomes of the order of the bulk population. 

\begin{figure}
  \centering
  \subfigure[]{
    \centering
    \includegraphics[width=0.45\textwidth]{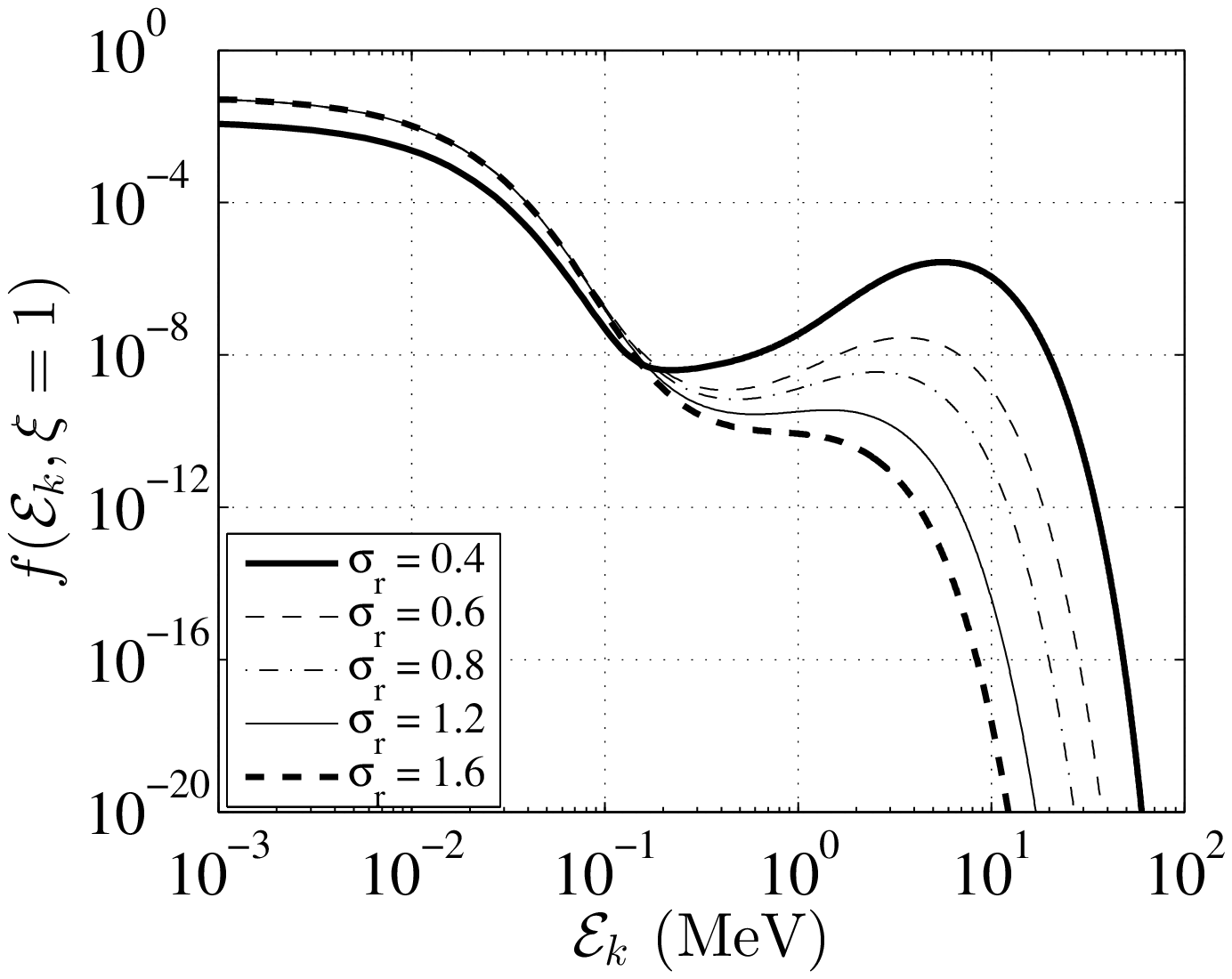}
    \label{fig:param_sigma-a}
  }
  \subfigure[]{
    \centering
    \includegraphics[width=0.45\textwidth]{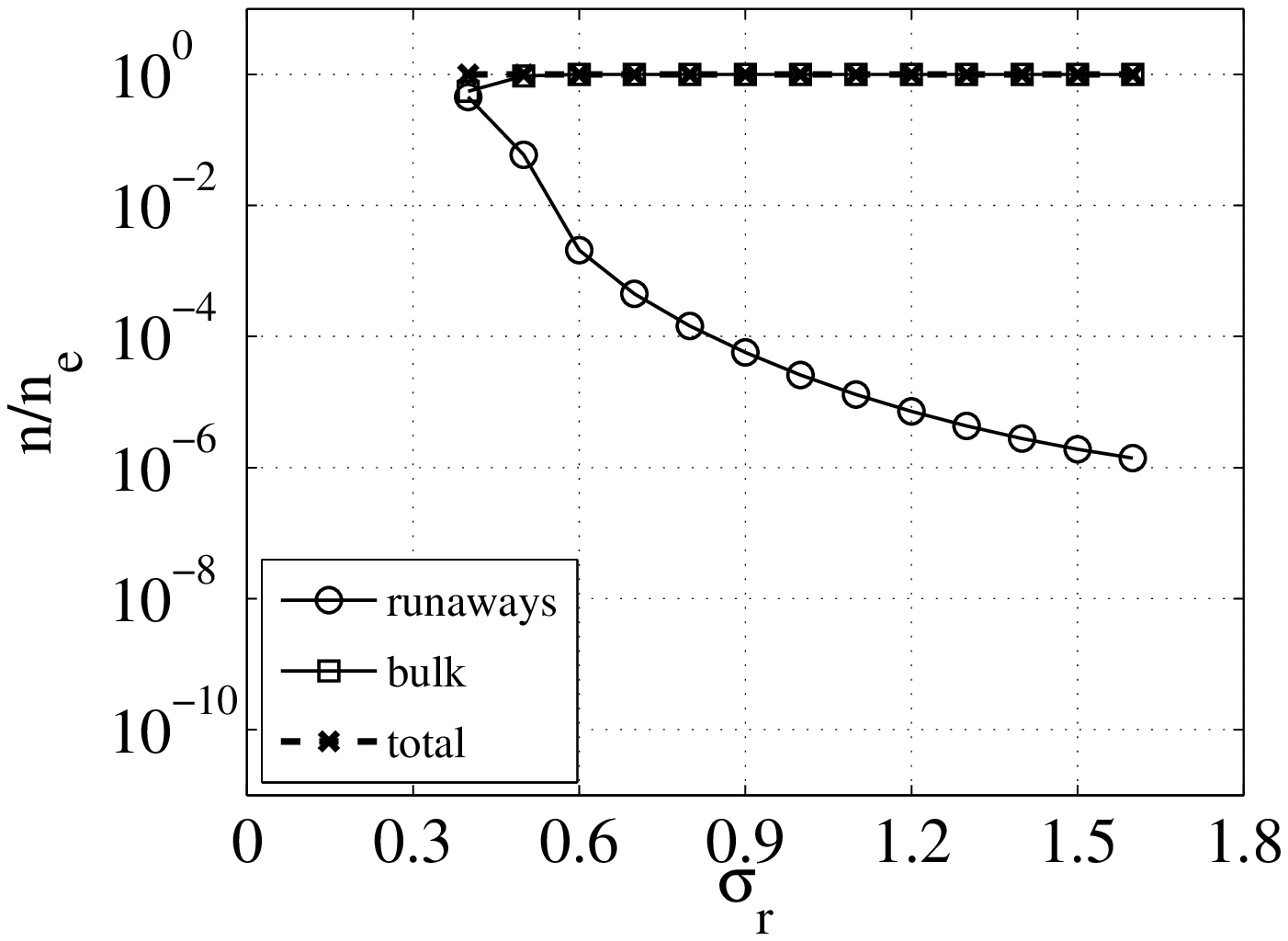}
    \label{fig:param_sigma-b}
  }
  \caption{(a) Electron distribution function in the parallel
    direction ($\xi=1$) as a function of the electron kinetic energy $\mathcal{E}_{k}$, and (b) fraction of electrons in the runaway region. The simulation
    results are plotted for various values of the synchrotron reaction
    force amplitude. Fixed relevant parameters are $Z_{\textrm{eff}}=1$,
    $\beta=0.1$, and $E_{\Vert}=3$. }
  \label{fig:param_sigma}
\end{figure}

In a second set of calculations, the electric field is fixed to $E_{\Vert}=3$,
while we vary the amplitude of the synchrotron radiation force, which
is proportional to $B^{2}$. The results are shown in Fig.~\ref{fig:param_sigma}.
As expected from (\ref{eq:bumploc}), we observe that the bump size and
the location of the bump maximum in energy both decrease
if $\sigma_{r}$ is increased, to the point where the bump disappears
if $\sigma_{r}$ is above a certain threshold. 

\begin{figure}
  \centering
  \subfigure[]{
    \centering
    \includegraphics[width=0.45\textwidth]{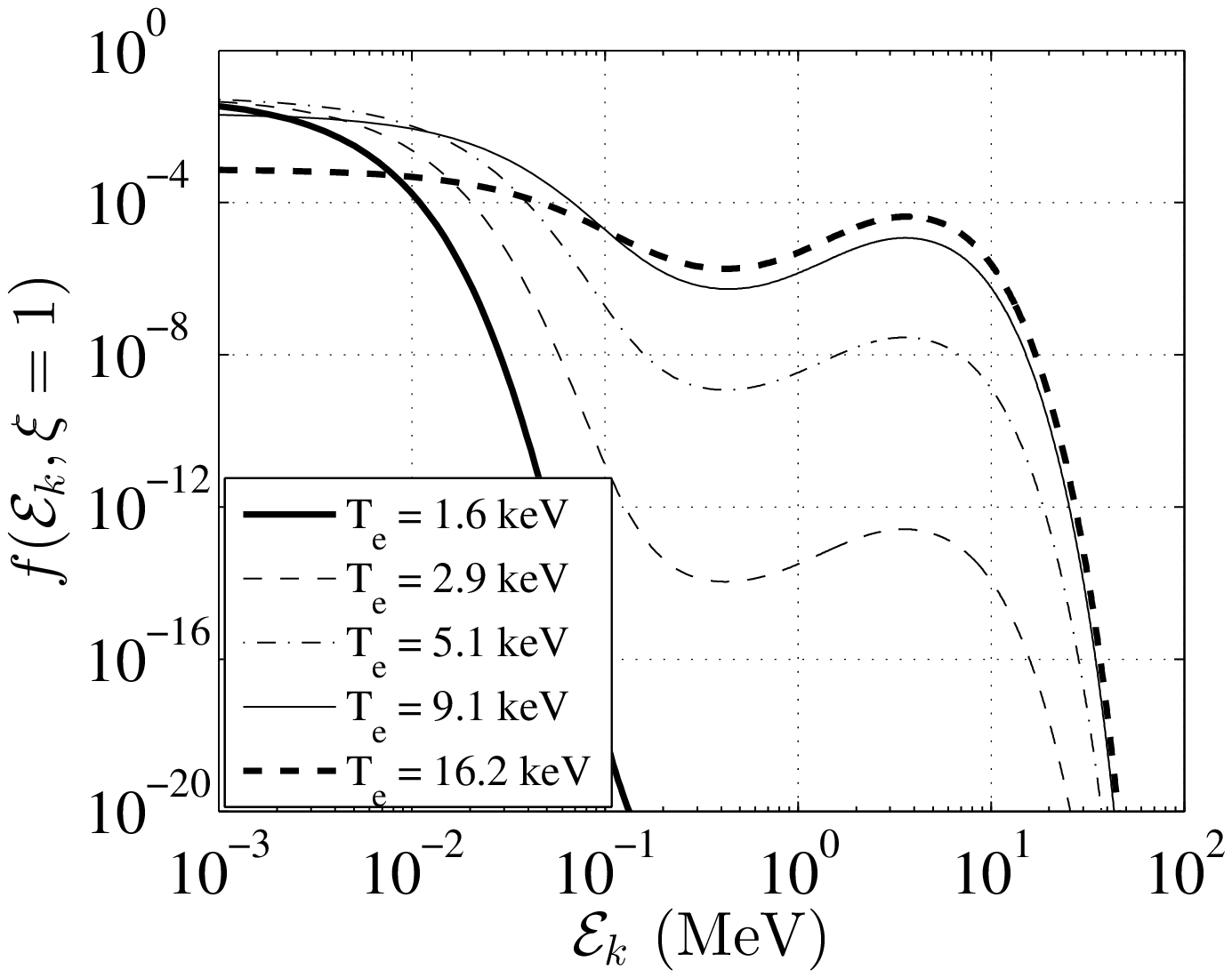}
    \label{fig:param_beta-a}
  }
  \subfigure[]{
    \centering
    \includegraphics[width=0.45\textwidth]{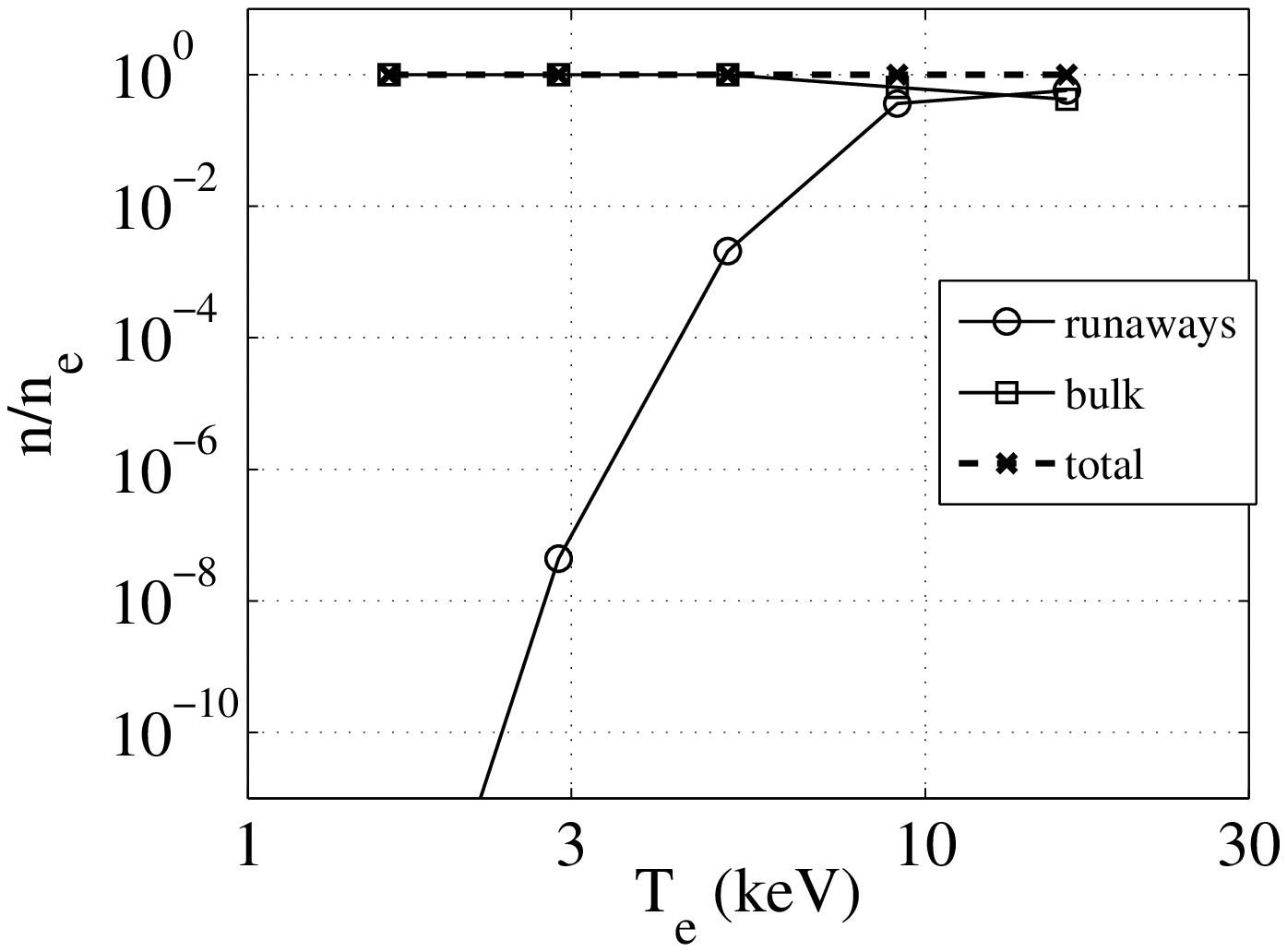}
    \label{fig:param_beta-b}
  }
  \caption{(a) Electron distribution function in the parallel
    direction ($\xi=1$) as a function of the electron kinetic energy $\mathcal{E}_{k}$, 
    and (b) fraction of electrons in the runaway region. The simulation
    results are plotted for various values of the electron temperature
    $T_{e}$. Fixed relevant parameters are $Z_{\textrm{eff}}=1$, $E_{\Vert}=3$,
    and $\sigma_{r}=0.6$. }
  \label{fig:param_beta}
\end{figure}

In a third set of calculations, the temperature is varied while the
normalized amplitudes of the electric field and synchrotron radiation
force are fixed. The results are shown in Fig.~\ref{fig:param_beta}.
We observe that the bump existence and energy are not affected by
the electron bulk temperature, which is again in accordance with the
analytical estimate (\ref{eq:bumploc}). However, the number of electrons
in the bump increases strongly with $T_{e}$, to the point where the
linearization of the collision operator fails for $T_{e}>10$ keV
with our choice of parameters. This dependence can be explained by
the Dreicer effect, which feeds the runaway population from the bulk
via collisional diffusion. As seen in Fig.~\ref{fig:fevol-b}, the
energy corresponding to the minimum between the bulk and
the runaway bump decreases with time, until it becomes of the order
of the critical energy. At this stage, the minimum coincides with the
point where forces balance, such that the collisional diffusion in
energy comes to a halt. In other words, the bump population increases
until the negative diffusive flux associated with the positive energy 
gradient of the bump is sufficient to compensate for the Dreicer flux. 
Since the latter strongly depends upon the bulk temperature, the bump population
evolves accordingly.

\begin{figure}
  \centering
  \subfigure[]{
    \centering
    \includegraphics[width=0.45\textwidth]{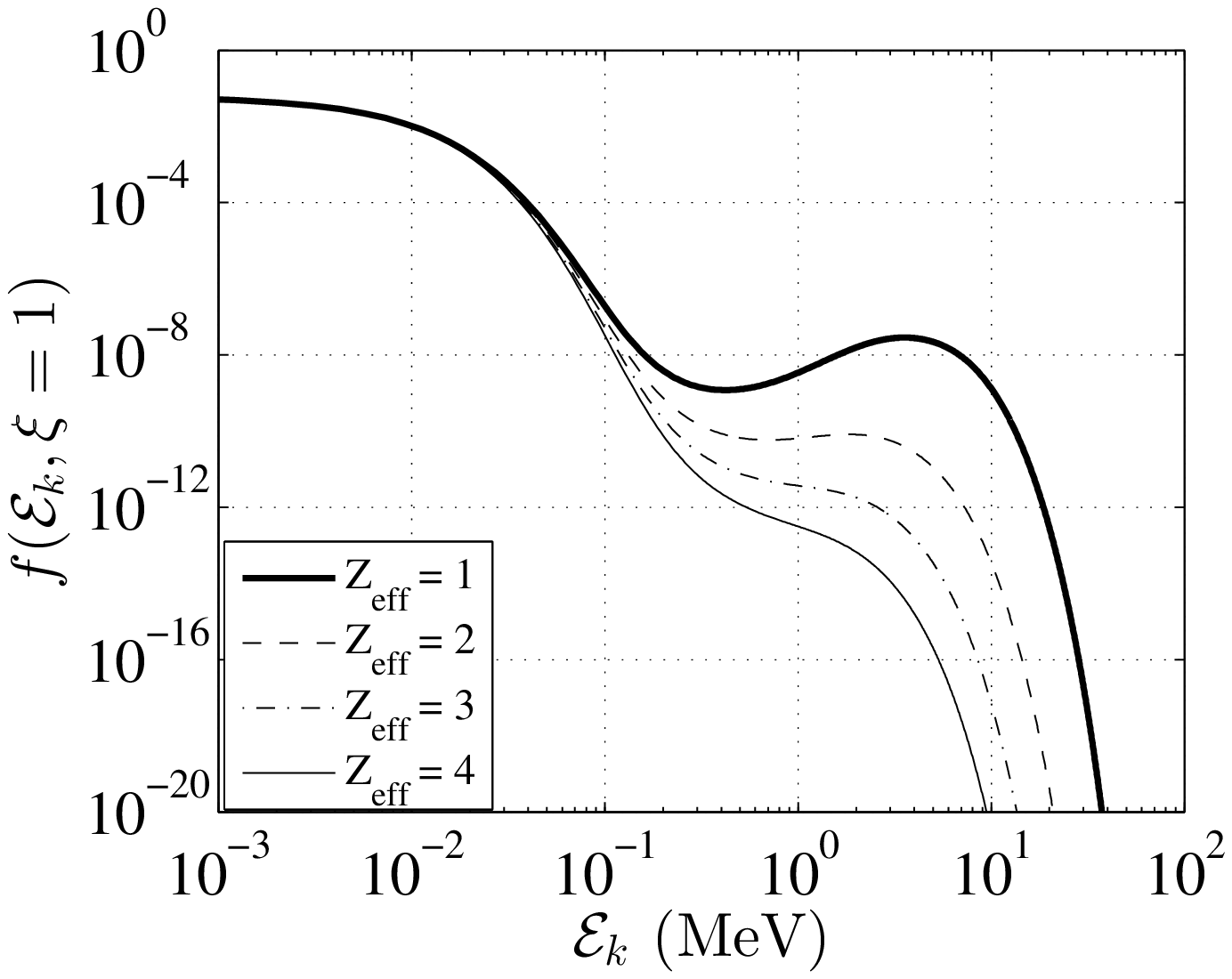}
    \label{fig:param_Zi-a}
  }
  \subfigure[]{
    \centering
    \includegraphics[width=0.45\textwidth]{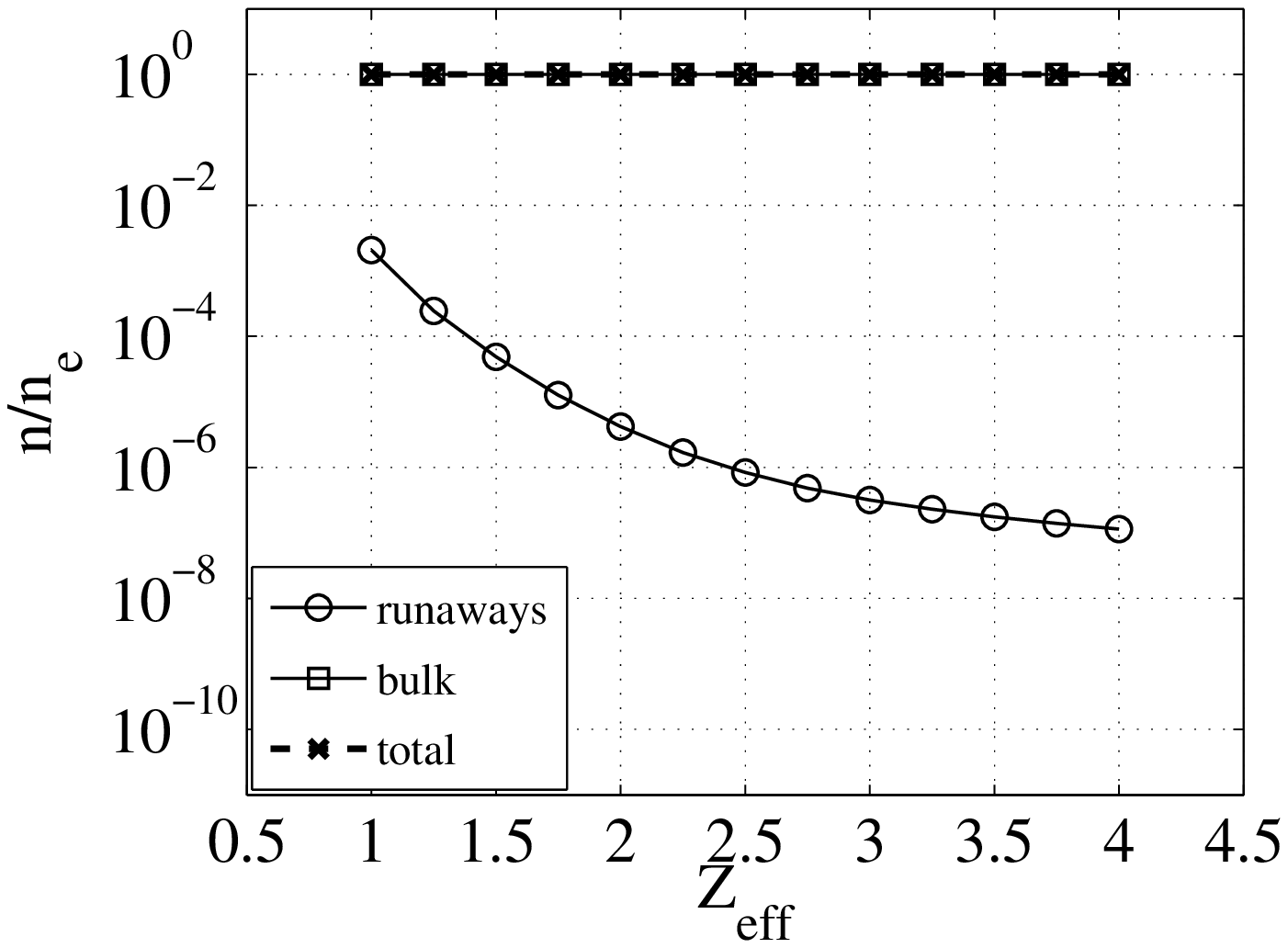}
    \label{fig:param_Zi-b}
  }
  \caption{(a) Electron distribution function in the parallel
    direction ($\xi=1$) as a function of the electron kinetic energy $\mathcal{E}_{k}$, 
    and (b) fraction of electrons in the runaway region. The simulation
    results are plotted for various values of the effective charge. Fixed
    relevant parameters are $\beta=0.1$, $E_{\Vert}=3$, and $\sigma_{r}=0.6$. }
  \label{fig:param_Zi}
\end{figure}

Finally, in a fourth set of calculations, the effective ion charge is varied
while all other relevant parameters are fixed. The results are shown
in Fig.~\ref{fig:param_Zi}. We observe that the bump size and energy
decrease with $Z_{\textrm{eff}}$, to the point where the bump disappears
for $Z_{\textrm{eff}}\geq3$ with this choice of parameters. The effect
of the ion effective charge is predicted by the estimate (\ref{eq:bumploc})
and understood via the role of pitch-angle scattering, which is proportional
to $1+Z_{\textrm{eff}}$. For a given perpendicular gradient in the
tail of the distribution function - which is determined by the parallel
force balance - pitch-angle scattering is the dominant mechanism to
extract electrons from the runaway region. The bump can exist only
if the perpendicular convection due to the synchrotron radiation force and collisional
drag dominates over pitch-angle scattering at the lower energies in the runaway region.

\section{Conclusions}\label{sec:Conclusions}

In this paper, the effect of the Abraham-Lorentz-Dirac force in reaction to the
synchrotron emission of runaway electrons is investigated for a homogeneous plasma. Whereas
a runaway region - with positive force balance - can still be
identified in the presence of the ALD force, the electron distribution
decreases with momentum at high energy and evolves towards a
steady-state solution.  This evolution is a result of the strong pitch-angle 
scattering associated with large gradients in perpendicular momentum. The 
distribution of electrons is limited to
energies well below the value for which the contribution from the
toroidal parallel motion to the ALD radiation reaction force becomes
significant in a tokamak plasma \cite{and01}, which justifies the
uniform plasma approximation to describe the runaway dynamics in the
plasma center.

If the electric force is large compared to the ALD force
(proportional to $B^{2}$) and the effective charge (which determines 
the rate of pitch-angle scattering), a bump centered
on the parallel momentum axis is formed in the steady-state electron distribution.
It results from the competition between the perpendicular convection
due to collisions and the ALD force, and pitch-angle scattering.
This bump encompasses almost the entire runaway electron population,
thus formally dividing the distribution into a bulk population and
a runaway beam. The steady-state population of electrons in the bump
is found to increase with the bulk electron temperature $T_{e}$.
The bump size and average energy increase with the electric field
amplitude $E_{\Vert}$, whereas they decrease with the amplitude of
the ALD radiation reaction force and the effective charge. 

We can summarize the effect of the ALD radiation reaction force on the
electron distribution in three points: first, in accordance with
experimental observations it limits the energy gained by runaway
electrons to the tens of MeV range; second, it increases the
perpendicular anisotropy of the electron distribution, which may give
rise to kinetic instabilities such as the EXEL or whistler waves
\cite{pokol,komar_pop,pokol2014}; third, it can lead to
the formation of a bump in the electron tail, which may give rise to
plasma-beam types of kinetic instabilities.

Large amplitude kinetic instabilities generated by the runaway population
could pump energy away from electrons and also affect their confinement.
Both effects could be beneficial when attempting to limit the threat
posed by runaway electrons in tokamaks. Quantifying the effect of
kinetic instabilities requires a quasilinear treatment of the kinetic
wave-particle interaction, which is beyond the scope of this paper.

More generally, it is interesting to note that any force with a magnitude
that increases with the particle energy could play a similar role as the ALD
radiation reaction force, resulting in a maximum energy limit for runaways and 
the possible formation of a bump in the energy distribution.

\begin{acknowledgments}
The work presented in this paper was done while J. D.~was invited to
work at Chalmers University under the Jubileum professorship award.
J. D.~would like to express his gratitude to T. F\"ul\"op, the eFT
group, and Chalmers University for this opportunity.
\end{acknowledgments}

\appendix

\section{Cylindrical representation\label{sub:Cylindrical-representation}}
The kinetic equation (\ref{eq:kineticexp}) can be expressed in the
$(p_{\Vert},p_{\bot})$ system, which yields 
\begin{equation}
\frac{\partial f}{\partial t}+\frac{1}{p_{\bot}}\frac{\partial}{\partial p_{\bot}}\left(p_{\bot}S_{\bot}\right)+\frac{\partial}{\partial p_{\Vert}}\left(S_{\Vert}\right)=0,\label{eq:kineticcyl}
\end{equation}
with the following expressions for the flux components 
\begin{equation}
\begin{aligned}
S_{\bot} = & -D_{\bot\Vert,\textrm{FP}}\frac{\partial f}{\partial p_{\Vert}}-D_{\bot\bot,\textrm{FP}}\frac{\partial f}{\partial p_{\bot}}\\
&+K_{\bot,\textrm{FP}}f+K_{\bot,\textrm{E}}f+K_{\bot,\textrm{ALD}}f,\\
S_{\Vert} = & -D_{\Vert\Vert,\textrm{FP}}\frac{\partial f}{\partial p_{\Vert}}-D_{\Vert\bot,\textrm{FP}}\frac{\partial f}{\partial p_{\bot}}\\
&+K_{\Vert,\textrm{FP}}f+K_{\Vert,\textrm{E}}f+K_{\Vert,\textrm{ALD}}f,
\end{aligned}
\label{eq:kineticfluxescyl}
\end{equation}
\begin{eqnarray*}
D_{\bot\bot,\textrm{FP}} & = & \frac{p_{\bot}^{2}}{p^{2}}A_{\textrm{FP}}+\frac{p_{\Vert}^{2}}{p^{2}}B_{\textrm{FP}},\\
D_{\bot\Vert,\textrm{FP}} & = & \frac{p_{\Vert}p_{\bot}}{p^{2}}\left(A_{\textrm{FP}}-B_{\textrm{FP}}\right),\\
D_{\Vert\bot,\textrm{FP}} & = & D_{\bot\Vert,\textrm{FP}},\\
D_{\Vert\Vert,\textrm{FP}} & = & \frac{p_{\Vert}^{2}}{p^{2}}A_{\textrm{FP}}+\frac{p_{\bot}^{2}}{p^{2}}B_{\textrm{FP}},
\end{eqnarray*}
\begin{eqnarray*}
K_{\bot,\textrm{FP}} & = & -\frac{p_{\bot}}{p}F_{\textrm{FP}},\\
K_{\Vert,\textrm{FP}} & = & -\frac{p_{\Vert}}{p}F_{\textrm{FP}},
\end{eqnarray*}
\begin{eqnarray*}
K_{\bot,\textrm{E}}^{\textrm{C}} & = & 0,\\
K_{\Vert,\textrm{E}}^{\textrm{C}} & = & E_{\Vert},
\end{eqnarray*}
\begin{eqnarray*}
K_{\bot,\textrm{ALD}}^{\textrm{C}} & = & -\sigma_{r}\frac{p_{\bot}}{\gamma}\left(1+p_{\bot}^{2}\right),\\
K_{\Vert,\textrm{ALD}}^{\textrm{C}} & = & -\sigma_{r}\frac{p_{\Vert}}{\gamma}p_{\bot}^{2}.
\end{eqnarray*}

\section{High-velocity limit\label{sub:High-velocity-limit}}
Properties of the electron distribution function are investigated
under the conditions that the thermal electron energy is much smaller
than the electron rest mass, namely $\beta\ll1$, and that the electric
field is larger than the critical field but much smaller than the
Dreicer field $E_{D}=\beta^{-2}$, meaning 
\begin{equation}
1<E_{\Vert}\ll\beta^{-2}
\end{equation}
This ordering implies that runaway electrons are located in the tail
of the distribution function, with a momentum $p\gg\beta$ where $\beta$
is the normalized thermal momentum. For such electrons it is appropriate to take the high velocity limit
of the collision operator, which yields 
\begin{align}
A_{\textrm{FP}}\left(p\right) &= \frac{\beta^{2}}{v^{3}},\\
F_{\textrm{FP}}\left(p\right) &=  \frac{1}{v^{2}},\\
B_{\textrm{FP}}\left(p\right) &= \frac{1+Z_{\textrm{eff}}}{2v}.
\end{align}

\section{ALD radiation reaction force for purely parallel motion\label{sub:parallel}}

In a non-uniform magnetic field, as is found in tokamaks, the field line
curvature affects the ALD radiation reaction force. Whereas this effect
is expected to be small compared to the contribution from the cyclotron
motion for particles with a significant magnetic moment, it could
play a role for particles with $\mathbf{p}_{\bot}\simeq 0$, for
which the contribution from the cyclotron motion (\ref{eq:K}) vanishes. The
combined effects of cyclotron motion and field curvature to the ALD
radiation reaction force have been evaluated in a previous work for
a purely toroidal magnetic field \cite{and01}. Whereas a self-consistent
calculation of the ALD radiation reaction force in a tokamak geometry
requires a proper guiding-center transformation \cite{hir14}, the importance
of the contribution from the field line curvature can be approximately
evaluated by considering the motion of a particle with $\mathbf{p}_{\bot}=0$.
The corresponding guiding center follows the field lines with a velocity
$\mathbf{v}=v_{\Vert}\hat{\boldsymbol{b}}$, such that the ALD radiation
reaction force (\ref{eq:reaction_force}) reduces to 
\begin{equation}
\mathbf{K}=\sigma_{r}\gamma^{2}v_{\Vert}^{3}\rho_{0}^{2}
\left[\hat{\boldsymbol{b}}\cdot\nabla\left(\hat{\boldsymbol{b}}\cdot\nabla\hat{\boldsymbol{b}}\right)
+\gamma^{2}v_{\Vert}\hat{\boldsymbol{b}}\cdot
\left[\hat{\boldsymbol{b}}\cdot\nabla\left(\hat{\boldsymbol{b}}\cdot\nabla\hat{\boldsymbol{b}}\right)\right]
\mathbf{v}\right],
\end{equation}
where the normalization of Sec. \ref{sub:gc} is used and with $\rho_{0}\equiv mc/(qB)$.
In a tokamak with major radius $R_{0}$ and circular concentric flux-surfaces
characterized by the local inverse aspect ratio $\varepsilon=r/R_{0}\ll1$,
$\mathbf{K}$ can be expressed as
\begin{equation}
\begin{aligned}
\mathbf{K}=&-\sigma_{r}\gamma\left(\frac{\rho_{0}}{R_{0}}\right)^{2}p_{\Vert}^{3}\left[\left(1-v_{\Vert}^{2}\left\{ 1-q^{2}\right\} \right)\varepsilon q^{-3}\hat{\boldsymbol{\theta}}\right.\\
&\left.+\left(1+2\varepsilon\cos\theta\left\{ q^{-2}-1\right\} \right)\hat{\boldsymbol{\phi}}+\mathcal{O}(\varepsilon^{2})\right],
\end{aligned}
\label{eq:Kcirc}
\end{equation}
where $\hat{\boldsymbol{\theta}}$ and $\hat{\boldsymbol{\phi}}$
denote the unit vectors in the poloidal and toroidal directions, respectively,
$\theta$ is the poloidal angle, and $q(\varepsilon)$ is the safety
factor. In the case of purely toroidal field lines ($q\rightarrow\infty$)
the results from Ref.\cite{and01} are retrieved.

The approximation $\varepsilon\ll1$ is valid near the
plasma center, where in addition we typically have $q\simeq1$, in
which case (\ref{eq:Kcirc}) becomes
\begin{equation}
\mathbf{K}=-\sigma_{r}\gamma\left(\frac{\rho_{0}}{R_{0}}\right)^{2}p_{\Vert}^{3}\left[\hat{\boldsymbol{b}}+\mathcal{O}(\varepsilon^{2})\right].\label{eq:kfinal}
\end{equation}

\bibliographystyle{unsrt}
\bibliography{bump}

\end{document}